\documentclass[12pt]{article}
\pdfoutput=1
\usepackage[pdftex]{graphicx}
\usepackage{amssymb}
\usepackage{amsmath}

\begin{document} 

\begin{titlepage}
	
	
	\vskip 2cm 
	\begin{center}
		\Large{{\bf Emergent spacetime\\\& Quantum Entanglement\\in Matrix theory}}
	\end{center}
	
	\vskip 2cm 
	\begin{center}
		{Vatche Sahakian, Yossathorn Tawabutr, and Cynthia Yan\footnote{\tt{sahakian@hmc.edu,\ ytawabutr@g.hmc.edu,\ xyan@g.hmc.edu}}}\\
	\end{center}
	\vskip 12pt 
	\centerline{\sl Harvey Mudd College} 
	\centerline{\sl Physics Department, 241 Platt Blvd.}
	\centerline{\sl Claremont CA 91711 USA}
	
	\vskip 1cm 
	\begin{abstract}
		In the context of the Bank-Fishler-Shenker-Susskind Matrix theory, we analyze a spherical membrane in light-cone M theory along with two asymptotically distant probes. In the appropriate energy regime, we find that the membrane behaves like a smeared Matrix black hole; and the spacetime geometry seen by the probes can become non-commutative even far away from regions of Planckian curvature. This arises from non-linear Matrix interactions where fast matrix modes lift a flat direction in the potential -- akin to the Paul trap phenomenon in atomic physics. In the regime where we do have a notion of emergent spacetime, we show that there is non-zero entanglement entropy between supergravity modes on the membrane and the probes. The computation can easily be generalized to other settings, and this can help develop a dictionary between entanglement entropy and local geometry -- similar to Ryu-Takayanagi but instead for asymptotically flat backgrounds. 
	\end{abstract}
\end{titlepage}

\newpage \setcounter{page}{1}

\section{Introduction and highlights}
\label{sub:intro}

Classical gravity is a phenomenon of geometrical origin, encoded in the curvature of spacetime. Quantum considerations however, whether in the setting of string theory or otherwise, suggest that the geometrical picture of gravity may be an effective long distance approximation scheme. It appears that at Planckian distances, a fundamental rethinking of the nature of gravity sets in. There have also been recent suggestions that the perception of gravity is entropic, arising from quantum entanglement~\cite{Ryu:2006bv}-\cite{Bao:2015bfa}. And subsequently, one talks about the concept of `emergent geometry': the idea that gravitational geometry is a collective phenomenon associated with underlying microscopic degrees of freedom.

In attempting to understand these ideas in a concrete computational setting, the Banks-Fishler-Shenker-Susskind (BFFS) Matrix model~\cite{Banks:1996vh} -- and its related cousin, the Berenstein-Maldacena-Nastase (BMN) system~\cite{Berenstein:2002jq} -- provide for a rich playground. They purport to describe quantum gravity in the full non-perturbative framework of light-cone M theory. The degrees of freedom are packaged into matrices that, in principle, encode geometrical gravity data at low enough energies. Spacetime curvature is then expected to arise from the collective dynamics of these matrix degrees of freedom. Unfortunately, the map between emergent geometry and matrix dynamics has proven to be a difficult one to unravel (but see recent progress in this direction~\cite{Anagnostopoulos:2007fw}-\cite{connor}).

A crude cartoon of Matrix theory dynamics goes as follows. The degrees of freedom, arranged in matrices, represent an interlinked complex web of membranes and fivebranes. At low energies, one can find settings where a hierarchy separates the different matrix degrees of freedom. Sub-blocs of the matrices, modes that remain light and slow, describe localized and widely separated lumps of energy; while other `off-diagonal' modes become heavy and frozen in their ground states -- heuristically corresponding to membranes/fivebranes stretched between the lumps. The effective dynamics of the lumps leads to the expected low energy supergravity dynamics, and hence a notion of emergent geometry. From this perspective, it is not surprising that a mechanism of entanglement across the degrees of freedom in sub-blocs of the matrices is key to the notion of emergent spacetime geometry. However, to our knowledge the role of quantum entanglement has not yet been explored in this context. In this work, our goal is to take the first steps in understanding how geometry may be encoded in Matrix theory degrees of freedom through a regime of heavy off-diagonal modes and through quantum entanglement of diagonal ones.

We consider a particularly simple setup in an attempt to make the otherwise challenging computation feasible. We will arrange a spherical membrane in light-cone M theory, stabilized externally so as to source a smooth static curved spacetime; and then we will add two probe supergravity particles a large distance away from the source. Using two probes instead of just one allows us to potentially capture a local invariant notion of gravity -- avoiding the possibility of missing out on emergent gravity due to the equivalence principle and background dependence. The configuration we consider is as expected also unstable in BFFS Matrix theory and the sphere needs to be externally stabilized\footnote{In BMN theory, there are metastable spherical configurations that can be used instead, but the BMN scenario is unnecessarily more complicated for the task at hand.}. Realizing the setup in matrices, we are immediately led to explore fluctuations of matrix modes that describes membranes stretched between sphere and probe. We first determine the relevant energy scale at which all off-diagonal matrix modes become heavy -- a criterion we expect to be a prerequisite for identifying emergent commutative geometry~\cite{Banks:1996vh,Berenstein:2008eg,Asplund:2012tg,Berenstein:2012ts,Berenstein:2014pma}. 
We find that the off-diagonal modes corresponding to the strings stretched between the two probes can become light far away from the center of the massive shell, {\em i.e.} far away from where spacetime curvature is Planck scale.

Using the configuration as a background scaffolding and integrating out heavy off-diagonal modes, we then focus on the effective dynamics of the diagonal fermionic modes. Zero modes of the fermionic degrees of freedom describe a system of qubits with a dense network of interactions. The qubit states map onto the eleven dimensional supergravity multiplet; hence, one is describing the interactions of supergravity modes in the given background. The setup for example has been used recently to demonstrate fast scrambling of supergravity modes in Matrix theory~\cite{Asplund:2011qj,Pramodh:2014jha}. We find that the effective Hamiltonian for the qubits includes direct couplings between qubits on the membrane and qubits on the probes -- when we focus on matrix dynamics with low enough energies to render the off-diagonal modes heavy and frozen.

We then proceed to finding the entanglement entropy between the sphere and probe qubits in the vacuum. Effectively, through the BFSS conjecture, we are computing the entanglement between supergravity modes on the sphere and on the probes. The computation can be carried out using expansions in several small parameters such as the ratio of the radius of the sphere to the sphere-probe distance. That is, we compute the entanglement entropy in the regime the probes are far away from the sphere.

The presentation is organized as follows. Section 2 gives an overview of the Matrix theory of interest, the M-theory perspective, and a sketch of the Matrix theory analysis. Section 3 presents the detailed computation in Matrix theory, the derivation of the effective Hamiltonian,  and the computation of entanglement entropy. Finally, Section 4 collects some concluding thoughts and directions for the future. Four appendices summarize technical details that arise in the main text. 

\section{The setup}

The BFSS theory is a 0+1 dimensional supersymmetric matrix theory (the dimensional reduction of the 10d super Yang-Mills (SYM)) describing the dynamics of D0 branes. For N D0 branes, the Lagrangian is given by~\cite{Banks:1996vh,Seiberg:1997ad,Bigatti:1997jy,Taylor:2001vb}
\begin{equation}\label{eq:bfssL}
	L = \mbox{Tr} \left[ 
	\frac{1}{2\,R}D_t{X}^i D_t{X}^i
	+ \frac{R}{4} \left[ X^i, X^j\right]^2
	+ i \Psi^{\dagger\, I}\cdot D_t{\Psi}^I
	- R \Psi^{\dagger\, I}\cdot \sigma^i\cdot \left[X^i,\Psi^I\right]
	\right]\ .
\end{equation}
The $X^i$'s are bosonic matrices, while the $\Psi^I_\alpha$'s are fermionic -- both are in the adjoint of $U(N)$. The full theory has $SO(9)$ symmetry, but for the purposes of the current work, we focus on a scenario where six of the nine target space directions are compactified and the corresponding excitations are frozen\footnote{One can think of the compact directions being of order the Planck scale so that associated excitations  would be much heavier than the energy scale we focus on. This is the case whether we use the dual $0+1$ dimensional SYM, or whether we T-dualize along the compact directions and consider a higher dimensional SYM theory.}. Hence, we are left with $SO(3)$ symmetry -- the index $i$ on $X^i$ runs from $1$ to $3$. Correspondingly, we also write the spinors  using $SU(4)\times SU(2)$ decomposition: on the $\Psi^I_\alpha$'s, I denotes the $SU(4)$ index and $\alpha$ is the $SU(2)$ label (not shown in the equation above). The $\sigma^i$'s are then the $2\times 2$ Pauli matrices. 
Using the static gauge, the covariant derivative $D_t$ becomes simply the time derivative -- at the cost of the constraint
\begin{equation}\label{eq:constrainteq}
	i [X^i, \Pi^i] + 2 \Psi^I\cdot \Psi^{\dagger\, I} = 0\ ,
\end{equation}
where the $\Pi^i$'s are the $X^i$'s canonical momenta.
The system is parameterized by
\begin{equation}\label{eq:RR11}
	R=\frac{g_s^{1/3}}{l_s} = \frac{R_{11}}{l_P^2}\ ,
\end{equation}
where $g_s$ is the string coupling, $l_s$ is the string length, $R_{11}=g_s l_s$ is the radius of the light-cone direction, and $l_P$ is the eleven dimensional Planck length. The Matrix theory decoupling regime corresponds to $R_{11},l_P\rightarrow 0$ while $R$ is held fixed.
In our conventions, $X$ and $\Psi$ are dimensionless, and time has unit of length; $X$ is then coordinate length measured in eleven dimensional Planck units.\footnote{To change to the conventions used in~\cite{Banks:1996vh}, write $X= g_s^{-1/3}\overline{X}$, $ t= g_s^{-2/3} \overline{t}$, and hence $H= g_s^{2/3} \overline{H}$ where the variables with bars correspond to the ones in~\cite{Banks:1996vh} in $l_s=1$ units.}.

The BFSS conjecture purports that this Lagrangian {fully} describes M-theory in the light-cone frame with $N$ units of light-cone momentum, $p_{\mbox{\tiny LC}}=N/R$, in the large $N$ limit. The matrix Hamiltonian is then identified with M-theory's light-cone energy
\begin{equation}
	H = \frac{m^2+p_i^2}{2\,p_{\mbox{\tiny LC}}}
\end{equation}
for a probe of mass $m$ and transverse momentum $p_i$.
While the BFSS conjecture was originally formulated in the large $N$ regime, the BFSS Matrix theory at finite $N$ is believed to describe discrete light-cone quantized (DLCQ) M-theory~\cite{Susskind:1997cw,Bigatti:1997jy}. However, to make our computation more tractable, we will assume that we are dealing with large matrices
\begin{equation}
	N\gg 1\ . 
\end{equation}

A massless supergravity particle with one unit of light-cone momentum is realized through the $N=1$ case: the bosonic part of the Hamiltonian reproduces the expected light-cone dispersion relation for a massless particle, while the zero modes of $\Psi^I_\alpha$'s give rise to the eleven dimensional supergravity multiplet -- the $256$ polarizations of the gravitons, gravitinos, and the 3-form gauge field~\cite{Banks:1996vh}. For more interesting setups, one starts with block diagonal configurations that break $U(N)\rightarrow U(N_1)\times U(N_2)\times \cdots$, and each matrix block can realize super-gravitons or membranes or fivebranes or black holes carrying different amounts of light-cone momenta~\cite{Banks:1996vh}\cite{Dijkgraaf:1997vv}-\cite{Banks:1997tn}. By developing the low energy quantum effective Hamiltonian for these blocks, one then reproduces eleven dimensional light-cone M-theory interactions at low supergravity energies. Furthermore, a proposed non-renormalization theorem sometimes allows a perturbative treatment in Matrix theory to still capture strongly coupled 
dynamics~\cite{Bigatti:1997jy}.

Consider the matrix configuration given by
\begin{equation}
	X^i=r L_i\ \ \ ,\ \ \ \Psi=0
\end{equation}
The $L_i$'s are $SU(2)$ matrices, satisfying $[L_i,L_j]=i\, \varepsilon_{ijk}L_k$, in an $N$ dimensional representation. This configuration represents an M-theory spherical membrane with N units of light-cone momentum -- sometimes called a non-commutative or fuzzy sphere -- of radius
\begin{equation}\label{eq:radius}
	\mathcal{R}=l_P\sqrt{\frac{\mbox{Tr} (X^i)^2}{N}} \simeq l_P\frac{r N}{2}\ \ \ \mbox{for large $N\gg 1$}\ .
\end{equation}
The matrix Hamiltonian then leads to the light-cone energy
\begin{equation}
	H = \frac{R}{2\,N}{M^2} \Rightarrow M=\frac{1}{l_P}\frac{N^2r^2}{2} = T_2\times 4\pi\, \mathcal{R}^2
\end{equation} 
where $T_2=1/2\,\pi\,l_P^3$ is the tension of the membrane. This configuration however is not a solution to the equations of motion. In particular, the potential $[X^i,X^j]^2$ appearing in the Hamiltonian provides for flat directions corresponding to mutually commuting matrices. Physically, this implies that it is energetically and entropically advantageous for this spherical membrane to eventually explode into widely separated super-gravitons. 

The BFSS conjecture has survived numerous checks (see for example~\cite{Banks:1996vh,Dijkgraaf:1997vv,Seiberg:1997ad}) and may be considered to be a background-dependent non-perturbative definition of light-cone M-theory. More recently, the BMN matrix model extended the setup to light-cone M theory in a plane wave background -- with the additional flux and curvature of the background geometry lifting the flat directions we alluded to above. The fuzzy sphere configuration then becomes a BPS classically stable configuration. These M-theory inspired matrix models can also be related to the AdS/CFT or gravitational holography conjecture~\cite{Maldacena:1997re,Witten:1998qj}. In practice however, the latter provides for a more precise dictionary between a gauge theory and string theory, while computations in the BFSS and BMN matrix theories quickly become technically very challenging and conceptually more difficult to interpret from the M-theory side. 

Paramount amongst the difficulties plaguing the BFSS/BMN settings is the challenge of understanding how the perception of spacetime is to emerge from matrix degrees of freedom. One natural approach is to identify the diagonal entries of the bosonic matrices $X^i$ as position labels (after all, they are related to the position of the underlying D0 branes). Implicit in this is that the usual notion of space geometry arises in the regime of heavy off-diagonal matrix modes. When matrix diagonal modes (or D0 brane positions) are widely separated, the off-diagonal matrix modes become heavy and frozen, leading to an effective dynamics for the diagonals that reproduces the supergravity interactions. Hence, in Matrix theory language, it seems the key to emergent gravitational geometry -- that is, the encoding of spacetime curvature information into matrix degrees of freedom -- lies in the interplay between the matrix eigenvalues {\em and heavy off-diagonal modes}. In the dual M-theory language, off-diagonal matrix degrees of freedom correspond to membranes stretched between the gravitating parts of the system. Presumably, it is then such a network of stretched M-theory membranes that underlies -- in the right low energy limit -- the perception of an emergent curved space. This may appear like an unusual perspective on emergent geometry, yet it syncs well with a seemingly independent line of thought that has recently risen in various other contexts: the concept of geometry or gravity emergent from quantum entanglement~\cite{Ryu:2006bv}-\cite{Czech:2014ppa}. A network of stretched membranes as represented by off-diagonal modes of matrices provides for a natural mechanism for entangling supergravity modes. Hence, in the right setting, we may perhaps expect to read off spacetime geometry data by looking at entanglement entropy in Matrix theory.

It is worthwhile noting that there have been several other different yet related approaches to the problem of emergent geometry in Matrix theory. In~\cite{Asplund:2012tg,Riggins:2012qt}, the focus has been on the bosonic dynamics of the matrix degrees of freedom. The system is highly non-linear and known to be chaotic~\cite{Maldacena:2015waa,Gur-Ari:2015rcq} and the idea here is that geometry emerges once one averages over the complex chaotic evolution resulting from couplings to off-diagonal matrix modes. In the commuting matrix regime, methods from random matrix theory (see for example~\cite{randommatrix} for a review) can be employed to extract statistical information about the eigenvalue distribution and the corresponding geometry. In~\cite{Berenstein:2013tya}, the role of the fermionic degrees were also considered in decoding geometry from matrices in the context of matrix black holes. Beyond the details, our approach also differs conceptually from previous attempts in that it focuses on a key new quantity -- the entanglement entropy of the fermionic matrix degrees of freedom. 

\subsection{M theory perspective}

\begin{figure}
	\begin{center}
		\includegraphics[width=5.5in]{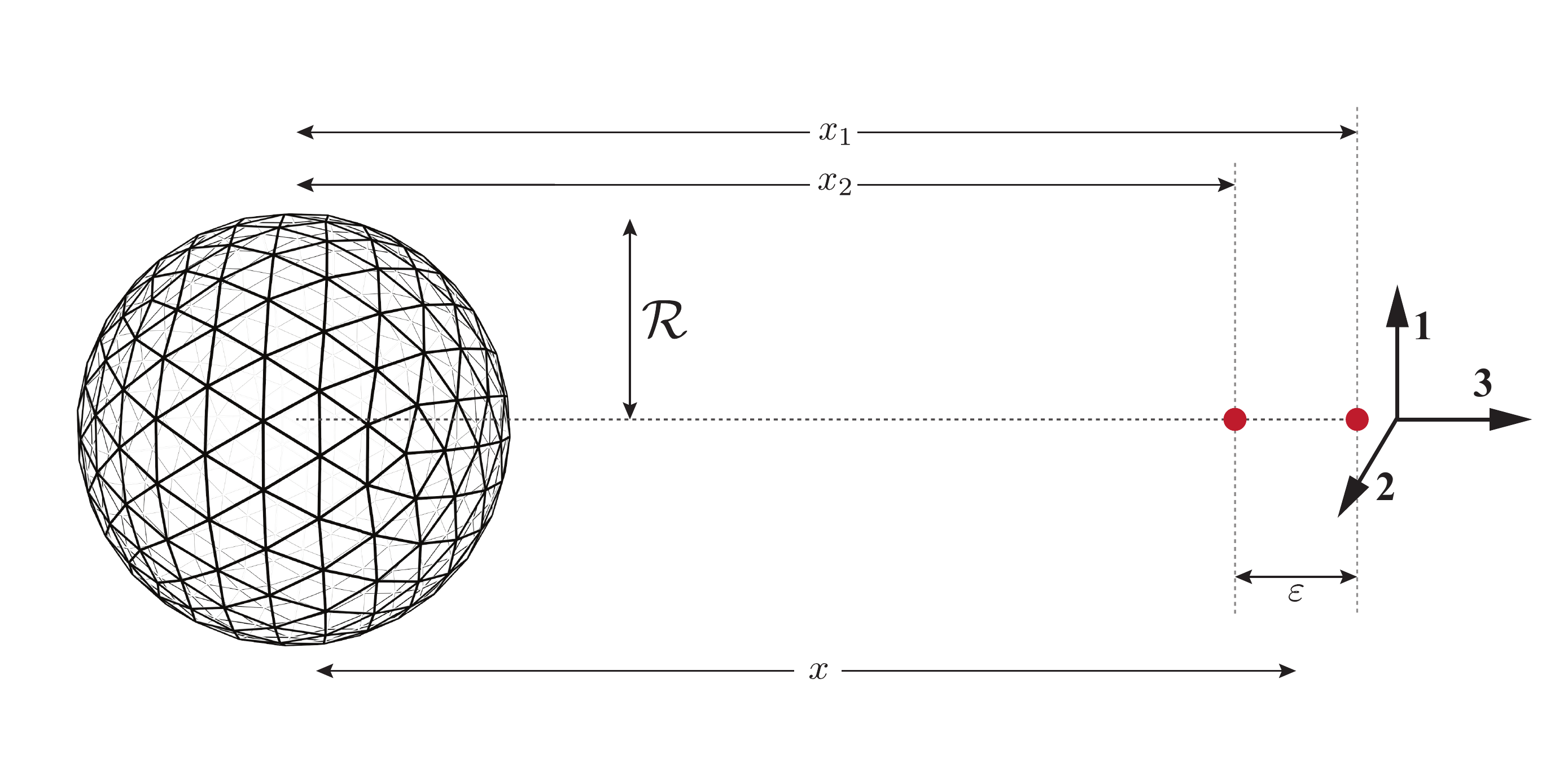}
	\end{center}
	\caption{A spherical membrane of fixed radius $\mathcal{R}$ and $N$ units of light-cone momentum, and two probes far away and colinear with it. $x_k=r_k/l_P$ with $k=1,2$ are the locations of the probes in eleven dimensional Planck units, and we define $\varepsilon = x_1-x_2$ while $x=(x_1+x_2)/2$.}\label{fig:setup}
\end{figure}

Figure~\ref{fig:setup} shows a cartoon of our setup. We have compactified to light-cone M-theory with three target space dimensions in addition to the light-cone direction; hence, six space directions are compact -- with size of order the eleven dimensional Planck scale $l_P$ -- while the seventh direction is compactified on a circle of size $R_{11}$ corresponding to the light-cone direction. Two probes with one unit of light-cone momentum each are positioned at a distance $x_1$ and $x_2$ away from a spherical membrane brane. The latter carries $N\gg 1$ units of light-cone momentum. And we choose $x_1>x_2$. We write $x_1$, $x_2$, $x=(x_1+x_2)/2$, and $\varepsilon = x_1-x_2>0$ as dimensionless distances, measured in eleven dimensional Planck units. The probes are arranged along the $3$-axis. 
The radius of the shell $\mathcal{R}$ is fixed and the shell's mass scales as
\begin{equation}\label{eq:massrad}
	M \sim \mathcal{R}^2/l_P^3 \gg l_P
\end{equation} 
where the tension of the membrane is $\sim l_P^{-3}$. As time evolves, the shell would collapse and disintegrate. Hence, we externally pin down the shell at the fixed radius $\mathcal{R}$. The probes are arranged a distance $x$ much greater than the size of the shell
\begin{equation}\label{eq:xcond}
	x \gg \frac{\mathcal{R}}{l_P} \gg 1\ ,
\end{equation}
while being close to each other
\begin{equation}
	\frac{\varepsilon}{x} \ll 1
\end{equation}
The Matrix theory regime, where light-cone M-theory can be mapped onto Matrix quantum mechanics, corresponds to\footnote{Note that this is a decoupling regime~\cite{Banks:1996vh,Sahakian:1999gj}: the Matrix theory-M theory correspondence is indeed a decoupling correspondence much like the AdS/CFT is.}
\begin{equation}\label{eq:matrixregime}
	l_P, R_{11}\rightarrow 0\ \ \ \mbox{while}\ \ \ \frac{R_{11}}{l_P^2} = \mbox{ fixed, }\ \frac{\mathcal{R}}{l_P},x, \varepsilon = \mbox{fixed}\ .
\end{equation}

\begin{figure}
	\begin{center}
		\includegraphics[width=5.0in]{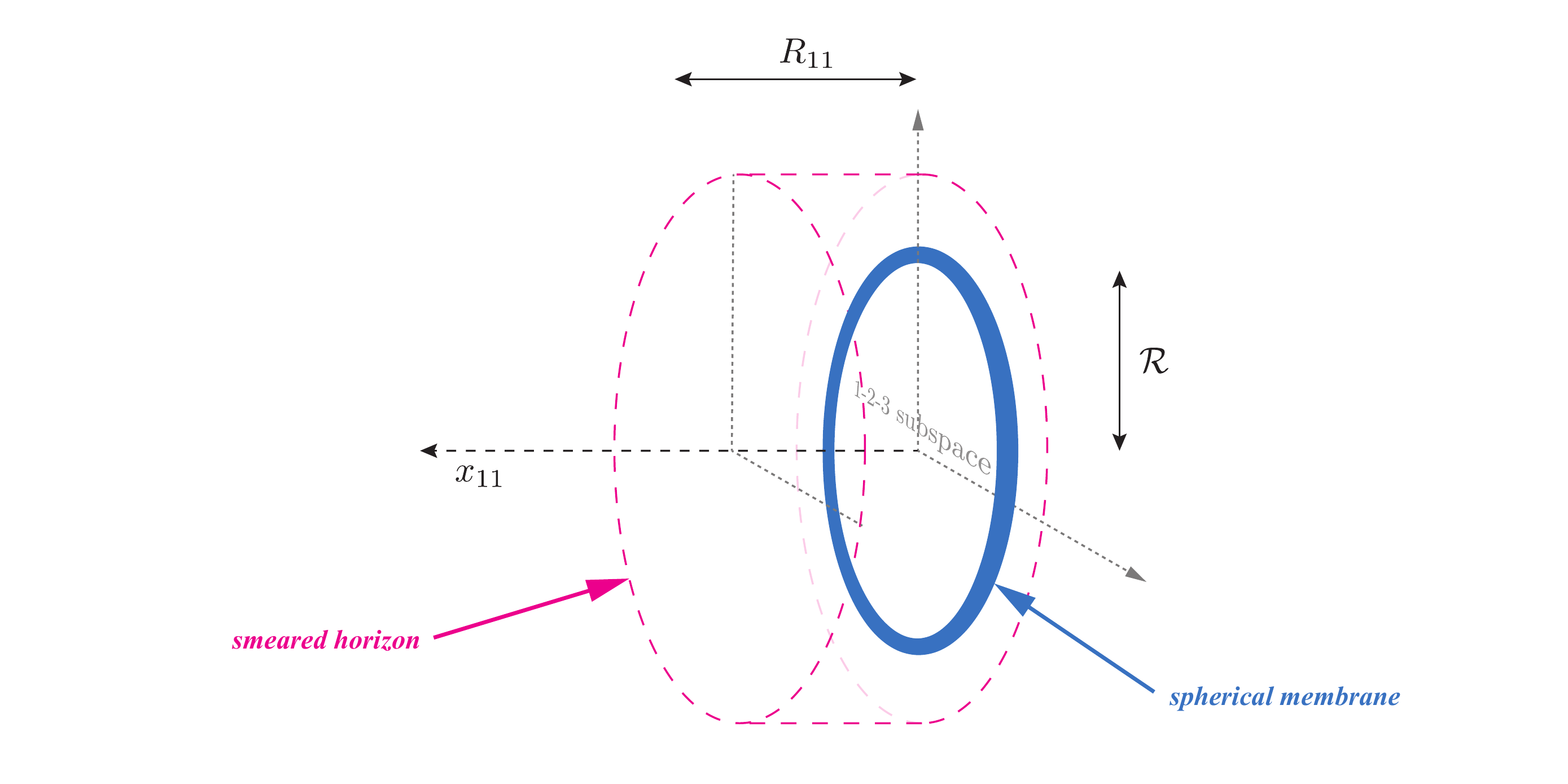}
	\end{center}
	\caption{A cartoon of the smeared light-cone black hole resulting from a heavy spherical membrane. A two dimensional cross section of the 1-2-3 subspace is shown.}\label{fig:smeared}
\end{figure}

For a large enough radius $\mathcal{R}\gg l_P$ for the spherical membrane, the object is massive enough to have a macroscopic horizon. Since we are in five dimensional (eleven minus the six Planck-size dimensions) M-theory, the horizon of the large membrane has radius
\begin{equation}
	r_0 \simeq \sqrt{G_5 M} \simeq \sqrt{l_p^3\, \mathcal{R}^2/l_P^3} = \mathcal{R}
\end{equation}
where $G_5$ is the five-dimensional Gravitational constant scaling as $l_P^3$. This is interesting, suggesting that --  because of the mass-radius relation of a membrane given by~(\ref{eq:massrad}) -- the horizon of the spherical membrane would be of order the size of the sphere. However, we are really dealing with a black hole in the light-cone frame, and hence we may expect that the horizon would be contracted from the large boost in the light-cone direction. As was demonstrated in~\cite{Horowitz:1997fr}, this is not the case; instead, the radius of the light-cone direction is expanded near the horizon by the boost factor
\begin{equation}\label{eq:boost}
	e^\alpha \sim \frac{N}{M\, R_{11}}
\end{equation}   
where $N/R_{11}$ is the light-cone momentum of the boosted hole and $M$ is its mass. This means that the light-cone radius expands as
\begin{equation}
	R_{11}\rightarrow \frac{N}{M} = \frac{N\,l_P^3}{\mathcal{R}^2}\ .
\end{equation}
In our case of a spherical membrane in light-cone M-theory, the ratio of the horizon of the object to $R_{11}$ becomes
\begin{equation}\label{eq:r0R11}
	\frac{r_0}{R_{11}} \rightarrow \frac{M\,r_0}{N} \simeq \frac{(M\,l_P)^{3/2}}{N} \simeq \frac{\mathcal{R}^{3}}{l_P^3\,N} \ .
\end{equation}
This is the volume of the spherical membrane in Planck units divided by $N$ and remains finite in the Matrix decoupling regime~(\ref{eq:matrixregime}). We will later see that our Matrix theory computation requires that this ratio is greater than one, 
which implies that the boosted spherical membrane we need to consider is in fact a {\em smeared} black hole~\cite{Martinec:1998ja}, smeared in the light-cone direction instead of being localized in all $4$ space direction. Figure~\ref{fig:smeared} shows a cartoon of this setup. 
We can then compute the horizon area as
\begin{equation}
	\mbox{Area} \sim \frac{N}{M} \mathcal{R}^2 \sim N\, l_P^3 \sim N\, G_5
\end{equation}
reproducing the expected entropy $S\sim N$ for a Matrix black hole~\cite{Horowitz:1997fr,Banks:1997hz,Banks:1997tn}. Hence, we have found that a large spherical membrane in light-cone M-theory in the proper regime acts as a smeared black hole whose horizon is located at around the membrane surface and whose entropy scales as expected with the horizon area. These conclusions are very much in tune with suggestions in~\cite{Horowitz:1997fr,Banks:1997hz,Banks:1997tn} about how one would want to model a black hole from Matrix theory's perspective.

Next, let us focus on the two probes which are arranged very far away from this spherical membrane. We are interested in the spacetime geometry seen by these two probes in low energy M-theory or eleven dimensional supergravity. In addition to the asymptotic form of a curved spacetime that we expect the probes to experience, the spherical membrane sources the supergravity 3-form gauge field. The net monopole membrane charge of the setup is zero, however we do expect higher multipole contributions. For example, as one looks towards the spherical membrane from any direction far away, one can approximate the spherical membrane by a dipole -- corresponding to equal but opposite membrane charges associated with two hemispheres oriented back-to-back. We want to put the probes far away enough that the effects of the membrane charge are asymptotically negligible as they are expected to fall off with distance faster than the influence of gravitons. We can capture this through a simple toy model of the setup, the details of which do not matter. Instead, the model will serve (a) to demonstrate that gauge field effects would be sub-leading, and (b) to read off the expected local curvature scale that the probes would see. We start by arranging a dipole from two five dimensional Reissner-Nordstr\"{o}m black holes of opposite charge. The separation between them would correspond to the scale $\mathcal{R}$ in our membrane problem. The five dimensional Reissner-Nordstr\"{o}m metric for a single black hole looks like
\begin{equation}\label{eq:RNmetric}
	ds^2 = -f(r) dt^2 + \frac{dr^2}{f(r)} + r^2 d\Omega_3^2
\end{equation}
with 
\begin{equation}
	f(r) = 1 - \frac{r_0^2}{r^2} + \frac{Q}{r^4}
\end{equation}
where 
\begin{equation}
	r_0^2 = \frac{8\, G_5 M}{3}\ .
\end{equation}
We immediately see that the monopole charge contribution to the metric dies off faster than the gravitational part by two powers of $r$. Furthermore, combining in the harmonic function $f(r)$ two close-by black holes with opposite charges, the dipole effect is even smaller by one more power of distance. Hence, for probes very far away, the leading physics is dictated by gravitons alone and we can write the metric~(\ref{eq:RNmetric}) with
\begin{equation}
	f(r) \simeq 1 - \frac{r_0^2}{r^2}\ ,
\end{equation}
{\em i.e.} leading to the five dimensional Schwarzschild metric. This black hole must be boosted to the light-cone frame, and this requires a large boost given by~(\ref{eq:boost}). To implement this, we introduce isotropic coordinates~\cite{Horowitz:1997fr,Sahakian:1999gj} by writing 
\begin{equation}
	{r}^2 = \frac{\rho^2}{4} \left(1+\frac{r_0^2}{\rho^2}\right)^2
\end{equation}
where $\rho^2 = x_{11}^2+\overline{r}^2$, 
which puts the metric in the form
\begin{equation}
	ds^2 = - \frac{H_-^2}{H_+^2} dt^2 + H_+^2 \left(dx_{11}^2+d\overline{r}^2 +\overline{r}^2 d\Omega_2^2\right)\ ,
\end{equation}
where we define
\begin{equation}
	H_\pm \equiv 1\pm\frac{r_0^2}{\rho^2}\ .
\end{equation}
We can now compactify and boost in the $x_{11}$ direction using~(\ref{eq:boost}) and we obtain the metric
\begin{equation}\label{eq:boostedmetric}
	ds^2 = \frac{1}{2} \left(H_+^2+ \frac{H_-^2}{H_+^2}\right) dx_+ dx_- + \frac{N^2}{M^2 R_{11}^2} \left(H_+^2- \frac{H_-^2}{H_+^2}\right) dx_-^2 + H_+^2 \left(d\overline{r}^2 +\overline{r}^2 d\Omega_2^2\right)
\end{equation}
where we now have
\begin{equation}
	H_\pm = 1 \pm \sum_n \frac{r_0^2}{\overline{r}^2+(N^2/M^2 R_{11}^2) (x_- + 2\pi n R_{11})^2}
\end{equation}
and $x_\pm = t \pm x_{11}$. In the Matrix theory limit~(\ref{eq:matrixregime}), $\overline{r} \sim l_P$ is large compared to $R_{11} \sim l_P^2$, we have effectively a four spacetime dimensions and the sum over images can be performed by integration leading to
\begin{equation}\label{eq:harmonics}
	H_\pm = 1 \pm \frac{M\, r_0^2}{2\,N\,\overline{r}}\ .
\end{equation}
We have a smeared black hole -- smeared in the $x_{11}$-direction. We will confirm this on the Matrix theory side later.
Note that the ratio $M\,r_0^2/2\,N\,\overline{r}$ remains finite in the Matrix theory decoupling limit~(\ref{eq:matrixregime}).
Hence, at very large distances from the setup, the probes will see at leading order the metric~(\ref{eq:boostedmetric}) with harmonic functions~(\ref{eq:harmonics}). Expanding for large $\overline{r}$, we can write asymptotically
\begin{equation}\label{eq:asmetric}
	ds^2 \rightarrow \left(1- \frac{4}{3} \frac{G_5 M^2}{N\,\overline{r}}\right) dx_+ dx_-
	+ \frac{8\,G_5 N}{R_{11}^2 \overline{r}} dx_-^2
	+ \left(1+ \frac{8}{3} \frac{G_5 M^2}{N\,\overline{r}}\right) \left(d\overline{r}^2 +\overline{r}^2 d\Omega_2^2\right)\ .
\end{equation}
Note that this result is model-independent: the asymptotic metric is determined by the symmetries and energy content of the source. Also, one can only use and trust this metric in the regime where
\begin{equation}\label{eq:metricexpansion}
	\frac{G_5 M^2}{N\,\overline{r}} \ll 1
\end{equation} 
since it is an asymptotic expansion valid only in such a domain. Investigating the form of this metric, we can easily see that assuring sub-Planckian curvature scales corresponds also to the condition given by~(\ref{eq:metricexpansion}). This does not however mean that the gemeotric depiction of spacetime is faltering when~(\ref{eq:metricexpansion}) is violated since the metric can only be trusted as an asymptotic form when~(\ref{eq:metricexpansion}) is satisfied.

It is also important to emphasize that this is not the usual near horizon regime arising in standard gravitational holography -- for good reason. If we were interested in the dynamics of the {\em spherical membrane}, we would naturally consider a $2+1$ dimensional non-commutative super Yang-Mills (NCSYM)~\cite{Maldacena:1999mh}, related to our D0 brane quantum mechanics, with an IR cutoff arising from the finite size of the sphere. At strong effective Yang-Mills coupling, this theory would have a geometric dual given by the near horizon geometry of an extremal black hole in M theory corresponding to our shell. Fluctuations in this spacetime would be mapped onto fluctuation in the NCSYM. In our setup, this would correspond to investigating particular low energy fluctuations in an $N\times N$ matrix sub-block of D0 branes -- which we are {\em not} interested in. We are instead insterested in the {\em asymptotically far away} geometry of the spherical membrane in light-cone M theory as seen by two massless probes -- which according to the Matrix theory conjecture should have a dual Matrix description. In Matrix theory language, this is captured by a $2\times 2$ probe matrix sub-block arranged a large distance from the membrane sub-block; the dynamical regime of interest is then such that the matrix sub-block corresponding to the shell acts as a set of non-dynamical background degrees of freedom that are fixed. The idea of emergent geometry in this setup then corresponds to the resulting effective matrix dynamics of the two probes `seeing' the asymptotic metric given by~(\ref{eq:asmetric}).

Having determined the spacetime geometry in the vicinity of the probes, it is interesting to determine the relation between energy and bulk distance, the so-called UV-IR relation, in the given background. Equating the first two terms of the metric~(\ref{eq:asmetric}), and writing $x_+\sim 1/E_{LC}$ and $x_-\sim 1/p_{LC}$, we quickly get
\begin{equation}\label{eq:uvir}
	E_{LC} \sim (R_{11} P_{LC}) \frac{R_{11}}{l_P^2} \frac{{x}}{N}
\end{equation}
where $x=\overline{r}/l_P$. This implies that large $x$ corresponds to larger light-cone energy: a probe needs more energy to overcome the gravitational pull of the shell. However, this energy scale is in a reference frame where spacetime has been infinitely boosted. It scales as $R_{11}/N = 1/p_{LC}$ indicating that this is an energy scale of a system carrying $N/R_{11}$ units of light-cone momentum. We are interested in the gravitational energy scale for a massless probe whose light-cone momentum is only $1/R_{11}$, located far away from a spherical membrane that is carrying $N$ units of light-cone momentum and acting as a fixed background. On general light-cone kinematic grounds~\cite{Bigatti:1997jy}, this interaction energy is expected to scale as
\begin{equation}
	E_{LC} \sim \frac{R_{11}}{l_P^2} U
\end{equation}
where $U$ is the `internal energy' of the probes -- in this case arising from a fixed background in which the probes move. Noting that $U$ must be dimensionless and finite in the Matrix theory limit~(\ref{eq:matrixregime}), we may guess that
\begin{equation}
	U\sim -\frac{M\times \frac{N}{M}}{x}\sim -\frac{N}{x}\ ,
\end{equation} 
where the $N/M$ factor comes from boosting the rest mass energy $M$ of the source to get it to $N$ units of light-cone momentum. Extending the result to two probes and the effect of their relative tidal gravitational force, we may write
\begin{equation}\label{eq:gravregime}
	E_{LC} \sim  -\frac{R\,N}{x} \left(1+\frac{\varepsilon}{x}\right)
\end{equation}
using~(\ref{eq:RR11}), and where $\varepsilon\ll x$. This would then be our estimate of what interaction energy scale to expect in a Matrix theory of two probes in the fixed background of a large Matrix shell\footnote{Note that the contribution to the light-cone energy of the probes coming from their own energies/momenta cancels: a massless probe with $E=p=1/{R_{11}}$ has $E_{LC} = E-p = 0$.}. 

\subsection{Matrix theory viewpoint and sketch of computation}

In Matrix theory language, we start with a static arrangement of $(N+2)\times (N+2)$ matrices of the form
\begin{equation}
	X^i, \Psi^I \rightarrow \left(\begin{array}{ccc}
	{\framebox[50pt][r]{\raisebox{0pt}[20pt][20pt]{\mbox{ $N\times N$ }}}}	&	
	{\framebox[11pt][r]{\raisebox{0pt}[20pt][20pt]{\mbox{ }}}} &
	{\framebox[11pt][r]{\raisebox{0pt}[20pt][20pt]{\mbox{ }}}}	 \\ 
	{\framebox[50pt][r]{\raisebox{0pt}[3pt][0pt]{\mbox{ }}}}	&
	{\framebox[11pt][r]{\raisebox{0pt}[3pt][0pt]{\mbox{ }}}}	&
	{\framebox[11pt][r]{\raisebox{0pt}[3pt][0pt]{\mbox{ }}}}	\\
	{\framebox[50pt][r]{\raisebox{0pt}[3pt][0pt]{\mbox{ }}}}	&
	{\framebox[11pt][r]{\raisebox{0pt}[3pt][0pt]{\mbox{ }}}}	&
	{\framebox[11pt][r]{\raisebox{0pt}[3pt][0pt]{\mbox{ }}}}
	\end{array}\right)\label{eq:blocks}\ .
\end{equation}
We then write

\begin{equation}
	X^i=\left(
	\begin{array}{ccc}
	 r L_i & 0 & 0 \\
	 0 & x^i_{1} & 0 \\
	 0 & 0 & x^i_{2} \\
	\end{array}
	\right)\ \ \ ,\ \ \ \Psi^I = 0\label{eq:vacuum}
\end{equation}
for $i=1,2,3$, with the $N\times N$ block representing the spherical membrane or shell, and the $x^i_{1}$ and $x^i_{2}$ representing the coordinates of the two probes. $X^i$ for $i=4,\cdots, 9$ are not involved: to assure a three dimensional target space, we imagine that the directions $i=4,\cdots, 9$ are compactified to Planck scale; excitations in these directions will be heavy compared to the energy scale we focus on\footnote{We can think of this in the T-dual picture as well where we would have a $6+1$ dimensional super Yang-Mills with frozen excitations along the worldvolume as the energy gap for such modes would be of order the Planck scale.}. Without loss of generality, we also arrange colinear probes and shell along the third axis so that $x^i_{1}=x^i_{2}=0$ for $i=1$ or $i=2$. Hence, we will eventually drop the $i$ index from $x^i_{k}$ and write instead $x_k\equiv x^3_{k}$ where $k=1,2$ labels the probes. Note also that $x^i_{k}$ and $r$ are dimensionless, written in Planck units. $r$ is a free parameter that allows us to tune the radius of the spherical shell for fixed $N$. It will also serve as an expansion parameter in the computation: for $r\gg 1$, we will be able to employ a perturbative approach in understanding the non-linear interactions of the theory.

As expected from the dual M-theory picture above, the configuration at hand is also unstable from Matrix theory's perspective. However, our problem statement is to fix the shell by external forces. In Matrix theory language, this means that the necessary terms are added to the action so that the $N\times N$ block in~(\ref{eq:blocks}) is stabilized.

One way we can quantify entanglement in this sytem is to compute the entanglement entropy between perturbations on the probes and on the shell. We write the bosonic perturbations as  
\begin{equation}
	\delta X^i=\left(
	\begin{array}{ccc}
	 \delta L_i & \delta x^i_{1} & \delta x^i_{2} \\
	 \delta x^{i\dagger}_{1} & \Delta x^{i}_{1} & \delta \xi^i\\
	 \delta x^{i\dagger}_{2} & {\delta \xi^{i\dagger}} & \Delta x^{i}_{2} \\
	\end{array}
	\right)\ .\label{eq:deltaXigeneral}
\end{equation}
Note that $\delta x^i_{k}$, with $k = 1,2$, is an $N$-component vector which can be written as 
\begin{equation}
	\delta x^i_{k} \rightarrow \delta x^i_{k\,m}
\end{equation}
where $m,n =-J, \cdots , J$, is a spin-$J$ representation index with $N=2J+1\simeq 2J$. The $\delta L_i$ corresponds to an $N\times N$ block of fluctuations of the shell. `Fixing the shell' in Matrix theory language would correspond to lifting the tachyonic modes amongst the $\delta L_i$'s.

And the fermionic fluctuations take the form
\begin{equation}\label{eq:fermions}
	\delta \Psi=\left(
	\begin{array}{ccc}
	 \Psi  & \delta \psi_1 & \delta \psi_2 \\
	 \overline{\delta \psi_1} & \psi_1 & \psi  \\
	 \overline{\delta \psi_2} & \overline{\psi} &
	   \psi_2 \\
	\end{array}
	\right)\ .
\end{equation}
In our conventions, $\delta \Psi$ is {\em not} Majorana, and a bar over a fermionic variable is {\em not} complex conjugation. We write the $\delta \psi_k$ and $\overline{\delta \psi}_k$ in component form
\begin{equation}
	\delta \psi_k, \overline{\delta \psi}_k \rightarrow \delta \psi^{\pm\, I}_{k\,m}, \overline{\delta \psi}^{\pm\, I}_{k\,m}
\end{equation}
The $\pm$ superscript refer to the eigenvalue of the mode under $\sigma^3$ of $SU(2)$, while $I=1,\ldots,4$ is the $SU(4)$ label; and $m$ is a spin-$J$ representation index as above. Similarly, the components of the other fermionic fluctuations can now be written as
\begin{equation}
	\psi, \overline{\psi} \rightarrow \psi^{\pm\, I}, \overline{\psi}^{\pm\, I}\ \ \ ,\ \ \ \psi_k \rightarrow \psi_k^{\pm\, I}
\end{equation}
and
\begin{equation}
	\Psi\rightarrow \Psi_{pq}^{\pm\, I}
\end{equation}
where in the last equation $p,q=1,\cdots N$ is a matrix row/column label. Henceforth, we will drop the $SU(4)$ index $I$ from all equations to reduce clutter: how to reinstate them will always be obvious from the required $SU(4)$ symmetry. 

We want to look at the {\em quantum entanglement} between supergravity fluctuations on the shell and the probes. To see how one can do this, 
we first quantize the fermionic fluctuations given their kinetic terms in~(\ref{eq:bfssL}), leading to
\begin{eqnarray}
	&&\left\{\psi^\pm_k , {\psi}_l^{\pm\dagger}\right\}=\delta_{kl}\ \ \ ,\ \ \ 
	\left\{\Psi^\pm_{pq} , {\Psi}_{p'q'}^{\pm\dagger}\right\}=\delta_{pp'} \delta_{qq'}\ , \nonumber \\
	&&\left\{\delta \psi^\pm_{k\,m} , {\delta \psi}_{l\,n}^{\pm \dagger}\right\}=\delta_{mn} \delta_{kl}\ \ \ ,\ \ \ \left\{\overline{\delta \psi}^\pm_{k\,m} , \overline{\delta \psi}_{l\,n}^{\pm \dagger}\right\}=\delta_{mn} \delta_{kl}\nonumber \\
	&& \left\{\psi^\pm , {\psi}^{\pm\dagger}\right\}=1\ \ \ ,\ \ \ \left\{\overline{\psi}^\pm , \overline{\psi}^{\pm\dagger}\right\}=1\ .
\end{eqnarray}
Hence, we have a system of $16\,N$ qubits in the $\delta \psi_k$'s, $8$ qubits in the $\psi$'s, and $16$ qubits in the $\psi_k$'s. Finally, the $\Psi$ correspond to $8\,N^2$ qubits associated with the shell. Note that this qubit system precisely maps onto polarizations of M theory's massless supergravity degrees of freedom. This was one of the original pieces of evidence in favor of the M theory-Matrix correspondence~\cite{Banks:1996vh}: $2^8=256$ states map onto the supergravity multiplet; and the ${N^2}$ comes from the decomposition of the multiplet amongst $N^2$ $U(N)$ discrete degrees of freedom. Hence, if we want to compute entanglement between M-theory supergravity modes on the shell and probes, we want to compute the quantum entanglement in this qubit system betweem the $\psi_k^\pm$ and the $\Psi_{pq}^\pm$.

Our task is then to derive the effective interaction with certain {\em fermionic modes as external legs}. In doing so, we will integrate out the heavy bosonic fluctuations from~(\ref{eq:deltaXigeneral}) at tree level. It can be easily seen that the $\delta L_i$'s do {\em not} contribute to this computation to leading tree level order: they don't appear as external legs and their are no quadratic fluctuations that are linear in the $\delta L_i$'s. 
Hence, for the purpose of capturing the leading non-zero contribution to the quantum entanglement between supergravity modes on the shell and probes, we can safely focus on the following bosonic perturbations only
\begin{equation}\label{eq:bosperturb}
	\delta X^i=\left(
	\begin{array}{ccc}
	 0 & \delta x^i_{1} & \delta x^i_{2} \\
	 \delta x_1^{i\dagger} & \Delta x^i_{1} & \delta \xi^i \\
	 \delta x_2^{i\dagger} & \delta \xi^{i\dagger} & \Delta x^i_{2} \\
	\end{array}
	\right)
\end{equation}
We emphasize that the full dynamics of the system involves all fluctuations. Yet since the leading contributions to entanglement between shell and probes comes from~(\ref{eq:bosperturb}), we choose to drop the irrelevant fluctuations from the outset. We must however assume that tachyonic $\delta L_i$'s have been lifted by external means. 

Under these conditions, let us first qualitatively sketch the computation of the effective Hamiltonian before getting into the details. We have from~(\ref{eq:bfssL}) schematically
\begin{eqnarray}
	H&\sim& \frac{1}{R}\delta \dot{x}^2 + R\,r^2\,\frac{x^2}{r^2}\,\delta x^2+ \frac{1}{R}\delta \dot{\xi}^2  - R\,r^2 \delta \xi^2 \frac{\varepsilon^2}{r^2} \nonumber \\
	&+& R\, \delta x \delta \psi \psi
	+ R\, r\, \frac{x}{r} \delta \psi \delta \psi + R\, r\, \psi \psi+ R\, r\, \frac{\varepsilon}{r}\psi \psi
	\label{eq:Hscaling}
\end{eqnarray}
where we write the bosonic perturbations $\delta x^i_{1}$ and $\delta x^i_{2}$ as simply $\delta x$, 
the bosonic perturbations $\delta \xi^{i}$ as $\delta \xi$,
the fermionic perturbations $\Psi$, $\psi_k$, $\psi$, and $\overline{\psi}$ corresponding to the shell and probes collectively as just $\psi$, 
the $\delta \psi_k$'s as $\delta \psi$, and both probe locations as $x$.  We also note that the last term in~(\ref{eq:Hscaling}) does not involve coupling between qubits on the probe and the shell, but couples instead probe qubits to probe qubits, and shell qubits to shell qubits. We have written the qualitative form of each term in the regime where $x/r\gg 1$. We have also dropped cubic and higher order bosonic terms as such terms come with higher powers of  $1/r$. Hence, to be able to control the computation, we require
\begin{equation}
	\frac{x}{r} \gg 1\ \ \ ,\ \ \ r\gg 1\ .
\end{equation}
Note that linear order terms vanish since fixing the background shell leads to a system that satisfies the equations of motion.

We expect that the notion of spacetime geometry arises when off-diagonal modes are heavy. From~(\ref{eq:Hscaling}), we can identify the energy scale for the $\delta x$'s and $\delta \psi$'s as $R\,x$; and the energy scale for the $\delta \xi$'s and off-diagonal $\psi$'s as $R\,\varepsilon$\footnote{This is as expected from string theory: the energy of strings stretched between D0 branes on the shell and probes is proportional to the distance $x$ between them, and the energy of strings stretched between the two probes is proportional to their separation $\varepsilon$.}. To be able to consider all off-diagonal modes as heavy, we would need to focus on the effective dynamics of the remaining qubits on the shell and probes with energy scale $E_{LC}$ satisfying
\begin{equation}\label{eq:Eregime}
	E_{LC} \ll R\,x\ \ \ \mbox{and}\ \ \ E_{LC}\ll R\,\varepsilon\ .
\end{equation}
From~(\ref{eq:xcond}), we also have
\begin{equation}
	x \gg r\,N \gg 1\ .
\end{equation}
Under these conditions, the off-diagonal modes are much heavier than the rest of the qubit dynamics. We are then left with an effective qubit Hamiltonian of the form 
\begin{equation}
	H^{eff} \sim  \frac{R}{r}\, \left<\delta x^2\right> \frac{r}{x}\psi\psi
\end{equation}
arising from integrating out the $\delta \psi$'s. The $\left<\delta x^2\right>$ is taken in the vacuum state of the heavy $\delta x$ oscillators. Hence, from~(\ref{eq:Hscaling}), it is given by
\begin{equation}
	\left<\delta x^2\right> \sim \frac{1}{r (x/r)}
\end{equation}
leading to an effective Hamiltonian 
\begin{equation}\label{eq:sketchscale}
	H^{eff} \sim 
	\frac{R}{r^2}\, \frac{1}{(x/r)^2}\psi\psi
\end{equation}
This generically would involve entanglement of qubits on the probes and the shell. We could then compute the entanglement entropy between supergravity modes on the shell and on the probes. Note also that the energy scale~(\ref{eq:sketchscale}) indeed conforms to~(\ref{eq:Eregime}) as needed provided $x\gg 1$ and $\varepsilon \gg 1$ which we naturally assume. 

We now proceed with carrying out the details of these computations. We will see that our general scaling analysis, while otherwise correctly capturing the physics, misses one interesting and important piece that arises from {\em quartic} couplings.

\section{Heavy modes and entanglement entropy}

To compute entanglement entropy in what will be a bilinear free qubit system, one needs to compute the correlation of any two qubits. A sketch of a general approach to such a computation is given in Appendix A. The key is to write an effective Hamiltonian for the qubits of interest, and find the correlation between the qubits in the state of interest by diagonalizing the resulting effective Hamiltonian. We start with the computation of the effective Hamiltonian by integrating heavy modes.

\subsection{The bosonic fluctuations}

Substituting~(\ref{eq:bosperturb}) into the first two terms of~(\ref{eq:bfssL}), it becomes convenient to rewrite the bosonic fluctuations in terms of $\delta X^\pm_{k\,m}$ and $\delta X_{k\,m}$
\begin{eqnarray}
	\left(\delta x^1_{k}\right)_m=\frac{1}{2} \left(\delta X^+_{k\,m}+\delta X^-_{k\,m}\right)\ \ \ &,&\ \ \ 
	\left(\delta x^2_{k}\right)_m=-\frac{1}{2} i \left(\delta X^+_{k\,m}-\delta X^-_{k\,m}\right)\nonumber \\
	\left(\delta x^3_{k}\right)_m&=&\frac{\delta X_{k\,m}}{\sqrt{2}}\label{eq:deltaxplusminus}
\end{eqnarray}
where $k=1,2$ labels the probe, and $m=-J,\cdots,J$ labels the spin in a spin-$J$ representation of the $\delta x$'s.
The kinetic term, written in the new variables, then looks like
\begin{equation}
	K = \frac{1}{2\,R} \left(\frac{d{\delta {X}}^{-\dagger}_{k}}{dt}\cdot \frac{d{\delta {X}^-}_{k}}{dt}
	+\frac{d{\delta {X}}^{+\dagger}_{k}}{dt}\cdot \frac{d{\delta {X}^+}_{k}}{dt}
	+\frac{d{\delta {X}}^\dagger_{k}}{dt}\cdot \frac{d{\delta {X}}_{k}}{dt} \right) + \frac{1}{R} \frac{d{\delta \xi}^{i\dagger}}{dt} \frac{d{\delta \xi^i}}{dt} \ ,
\end{equation}

The quartic potential exhibits the following pattern: writing all occurrences of the distances $x_k$ as $x_k/r$, all terms quadratic in the fluctuations scale as $r^2$, all cubic terms scale as $r$, and all quartic terms scale as $r^0$. Hence, to have a well-defined perturbative treatment in small fluctuations we need
\begin{equation}
	r \gg 1\ .
\end{equation}
This implies in turn that the ratio of horizon radius to the shell's radius given by~(\ref{eq:r0R11}) is
\begin{equation}
	\frac{r_0}{{R}_{11}} \rightarrow \frac{M\,r_0}{N} \simeq \frac{\mathcal{R}^3}{l_P^3 N} = r^3 N^2 \gg 1
\end{equation}
as mentioned earlier: the black hole corresponding to our setup is indeed smeared in the light-cone direction.
To quadratic order, the potential takes the form
\begin{eqnarray}
	V &=& R\,r^2\,\left(
	\begin{array}{ccc}
	 {\delta {X}}^{+\dagger}_{k\,m+1} & {\delta {X}}^{-\dagger}_{k\,m-1} & {\delta {X}}^\dagger_{k\,m} \\
	\end{array}
	\right)\cdot \mathcal{M}_k \cdot \left(
	\begin{array}{c}
	 \delta X^+_{k\,m+1} \\
	 \delta X^-_{k\,m-1} \\
	 \delta X_{k\,m} \\
	\end{array}
	\right) \nonumber \\
	&+& R\,r^2 \left(\frac{x_1}{r}-\frac{x_2}{r}\right)^2 \left(\delta\xi^{1\dagger}\delta\xi^1+\delta\xi^{2\dagger}\delta\xi^2\right) \nonumber \\
	\label{eq:Vmatfull}
\end{eqnarray}
where the mass matrix is\footnote{We have dropped from the potential a sub-leading term of the form $\kappa_{1m} \kappa_{2m}\delta X^{+\dagger}_{k\,m-1}\delta X^-_{k\,m+1}$ that otherwise makes the mass matrix diagonalization procedure considerable messier. As we will see, we will eventually need to keep only the leading order terms in this sector, terms that scale like $x^2$; and hence it will not be necessary to include this term in the problem at leading order in the computation.}
\begin{eqnarray}
	\mathcal{M}_k &=&
	\left(
	\begin{array}{ccc}
	 \frac{1}{2}\left(\frac{x_k}{r}\right)^2& 0 & 0 \\
	 0 & \frac{1}{2}\left(\frac{x_k}{r}\right)^2 & 0 \\
	 0 & 0 & 0 \\
	\end{array}
	\right)
+	\left(
	\begin{array}{ccc}
	 -(m+1)\frac{x_k}{r} & 0 & \kappa_{1m}\frac{x_k}{r} \\
	 0 & -(m-1)\frac{x_k}{r} & \kappa_{2m}\frac{x_k}{r} \\
	 \kappa_{1m}\frac{x_k}{r} & \kappa_{2m}\frac{x_k}{r} & 0 \\
	\end{array}
	\right) \nonumber \\
	&+& 
	\left(
	\begin{array}{ccc}
	 4\, J( J+1)+ \frac{1+m}{4} & -\kappa_{1m}\kappa_{2m} & -(m+1)\kappa_{1m} \\
	 -\kappa_{1m}\kappa_{2m} & 4\, J( J+1)+ \frac{1-m}{4} & -(m-1)\kappa_{2m} \\
	 -(m+1)\kappa_{1m} & -(m-1)\kappa_{2m} & 4\, J( J+1) \\
	\end{array}
	\right)
	\ .
\end{eqnarray}
Here, we have defined
\begin{equation}
	\kappa_{1m}=\frac{\sqrt{J (J+1)-m (m+1)}}{2 \sqrt{2}}\ \ \ ,\ \ \ 
	\kappa_{2m}=\frac{\sqrt{J (J+1)-m (m-1)}}{2 \sqrt{2}}\ ,
\end{equation}
We first perturbatively diagonalize this mass matrix, expanding in $r/x$. We find the three eigenvectors
\begin{eqnarray}
	&&V_{k\,1\,m} = \delta X^+_{k\,m+1} 
	- \frac{\kappa_{1m}\kappa_{2m}}{2\,(x_k/r)} \delta X^-_{k\,m-1} 
	+ \frac{2\,\kappa_{1m}}{(x_k/r)} \delta X_{k\,m} 
	- \left(\frac{\kappa_{1m}^2\kappa_{2m}^2}{8\,(x_k/r)^2} 
	+ \frac{2\,\kappa_{1m}^2}{(x_k/r)^2}\right) \delta X^+_{k\,m+1} \nonumber \\ 
	&&V_{k\,2\,m} = \delta X^-_{k\,m-1} 
	+ \frac{\kappa_{1m}\kappa_{2m}}{2\,(x_k/r)} \delta X^+_{k\,m+1}
	- \frac{\kappa_{1m}^2\kappa_{2m}^2}{8\,(x_k/r)^2} \delta X^-_{k\,m-1} 
	- \left(\frac{\kappa_{1m}\kappa_{2m}}{(x_k/r)^2}+
	\frac{2\,\kappa_{2m} (m-1)}{(x_k/r)^2}\right) \delta X_{k\,m}\nonumber \\ 
	&&V_{k\,3\,m} = \delta X_{k\,m}-\frac{2\,\kappa_{1m}}{(x_k/r)} \delta X^+_{k\,m+1} 
	+ \frac{2\,\kappa_{2m} (m-1)}{(x_k/r)^2} \delta X^-_{k\,m-1}
	+ \frac{2\,\kappa_{1m}}{(x_k/r)^2} \delta X_{k\,m}\ ,
\end{eqnarray}
with corresponding eigenvalues
\begin{eqnarray}
	\Lambda_{1\,m}&=&\frac{1}{2}\left(\frac{x_k}{r}\right)^2-(m+1)\frac{x_k}{r}+34\,\kappa_{1m}^2+\frac{1}{4}(m+1)(16\,m+1)\ , \nonumber \\
	\Lambda_{2\,m}&=&\frac{1}{2}\left(\frac{x_k}{r}\right)^2-(m-1)\frac{x_k}{r}+34\,\kappa_{2m}^2+\frac{1}{4}(m-1)(16\,m-1)\ , \nonumber \\
	\Lambda_{3\,m}&=& 4 J (J+1) - 2 \left(\kappa_{1m}^2+\kappa_{2m}^2\right)\ .
\end{eqnarray}
It is worthwhile noting several technical points:
\begin{itemize}
\item The diagonalization is computed perturbatively in $r/x \ll 1$. This requires a second order degenerate perturbation approach in the $\delta X^+_{k\,m+1}$ and $\delta X^-_{k\,m-1}$ sub-block following~\cite{gottfried}. The second order treatment is needed because the third eigenvalue would otherwise vanish and its first non-zero contribution comes from an $x$-independent term.
\item We find no contribution from fluctuations of the locations of the probes $\Delta x^i_{k}$ at quadratic order in small fluctuations.
\item Most of the terms coming at second order in perturbation theory do not at the end play a role in the leading order entanglement entropy computation. We present them here for a consistent second order treatment. 
\end{itemize}

We note that the first two eigenvalues correspond to a frequency scaling as $R\, x$ as expected from~(\ref{eq:Hscaling}) and hence are heavy in the regime~(\ref{eq:Eregime}). However, the third one scales as $R\, N$, arising from lighter fluctuations along the colinear direction of the probes with the shell. To assure that these modes are also heavy and hence we are in the regime of emergent spacetime, we need to require
\begin{equation}
	E_{LC} \ll R\, N\label{eq:ERN}
\end{equation}
in addition to~(\ref{eq:Eregime}).
Thus, the $\delta X$ modes are heavy and can eventually be frozen in the oscillator ground state giving them vacuum expectation value (vev)
\begin{equation}\label{eq:Vs}
	\left<\overline{V}_{k\,i\,m}{V}_{k\,i\,m}\right> = \frac{1}{r\,\sqrt{2\,\Lambda_{i\,m}}}
\end{equation}
for $i=1,2$, $k=1,2$, and $m=-J, \cdots J$. Note however that we will not freeze these oscillators until {\em after} we have integrated out the heavy off-diagonal fermionic modes. This is because there are couplings in the Hamiltonian linear in both the $\delta X$'s and the fermionic modes. The proper sequence of computational steps involves first integrating out the heavy fermionic modes generating new small terms quadratic in the bosonic off-diagonals, and then freezing the heavy bosonic modes in their vacuum states. Hence, after integrating out the fermionic modes, we can use
the following non-zero vevs for the $\delta X$'s that follow from~(\ref{eq:Vs})
\begin{eqnarray}
		\left<{\delta X}^{+\dagger}_{k\,m+1}{\delta X}^+_{k\,m+1}\right> & \simeq & \frac{1}{r\,(x_k/r)} +\left( \frac{m+1}{r\,(x_k/r)^2} -  \frac{2\,\kappa_{1m}}{C_m\,r\,(x_k/r)^2} \right)\ , \nonumber \\
		\left<{\delta X}^{-\dagger}_{k\,m-1}{\delta X}^-_{k\,m-1}\right> & \simeq & \frac{1}{r\,(x_k/r)} +\left( \frac{m-1}{r\,(x_k/r)^2}\right)\ , \nonumber \\
		\left<{\delta X}^\dagger_{k\,m}{\delta X}_{k\,m}\right> & \simeq & \frac{1}{2\,r\,C_m} +\left( \frac{2\,\kappa_{1m}}{C_m\,r\,(x_k/r)^2}\right) \ , \nonumber \\
		\left<{\delta X}^{+\dagger}_{k\,m+1}{\delta X}_{k\,m}\right> & \simeq & +\frac{\kappa_{1m}}{C_m\,r\,(x_k/r)} - \left( \frac{2\,\kappa_{1m}}{r\,(x_k/r)^2}\right)\ , \nonumber \\
		\left<{\delta X}^{-\dagger}_{k\,-\,m-1}{\delta X}_{k\,m}\right> & \simeq & 0 + \left( \frac{2\,(m-1) \kappa_{2m}}{2\,C_m\,r\,(x_k/r)^2}\right)\ , \nonumber \\
		\left<{\delta X}^{+\dagger}_{k\,+\,m+1}{\delta X}^-_{k\,m-1}\right> & \simeq & 0
		\ ,\label{eq:vevs1}
\end{eqnarray} 
where we have defined
\begin{equation}
	C_m^2\equiv 2\, J (J+1)-\kappa_{1m}^2-\kappa_{2m}^2\ .
\end{equation}
The parenthesized terms in~(\ref{eq:vevs1}) will not contribute to the final result and are included for the purposes of a consistent second order perturbation treatment. 

We now come back to~(\ref{eq:Vmatfull}) and the $\delta \xi^i$ terms. 
The frequency of the $\delta \xi^{1,2}$ modes scales as $R\,\varepsilon$. By~(\ref{eq:Eregime}), these are heavy and can be eventually frozen in the vacuum state giving
\begin{equation}
	\left<\delta \xi^{1\dagger} \delta \xi^1\right> = \left< \delta \xi^{2\dagger} \delta \xi^2 \right> = \frac{1}{r\,(x_1/r-x_2/r)}\ .
\end{equation}
However, the energy of the $\delta \xi^3$ mode vanishes at quadratic order. This is not due to any symmetry: we just need to look at higher order terms in the quartic potential of~(\ref{eq:bfssL}). 
The physics here is very similar to ion trapping with a Paul trap~\cite{paultrap} in electromagnetism. There, a tachyonic direction is lifted by modes that couple to it and that are much faster, {\em i.e.} visualize the spinning saddle model that renders the saddle point stable. In this case, we have a {\em flat} direction that is lifted by the faster $\delta X^{\pm}$ modes. To see this, we write {\em all} the $\delta \xi^3$ dependent terms of the quartic potential -- in the vacuum of the fast $\delta X^\pm$ oscillators; one gets
\begin{eqnarray}
	 & &\frac{R}{2} \left( \left<{\delta {X}}^{+\dagger}_{k\,m+1}{\delta {X}}^+_{k\,m+1}\right>+\left<{\delta {X}}^{-\dagger}_{k\,m-1}{\delta {X}}^-_{k\,m-1}\right>\right) \delta\xi^{3\dagger}\delta\xi^3 \nonumber \\
	 & & \simeq  \frac{R}{r} \frac{(x_1/r)+(x_2/r)}{x_1 x_2/r^2} \delta\xi^{3\dagger}\delta\xi^3 \label{eq:penning}
\end{eqnarray}
where we used~(\ref{eq:vevs1}) from above. This is similar to the Paul trap phenomenon where fast modes stabilize the effective dynamics of slower modes.
We see that the frequency of the $\delta \xi^3$ modes scales as $R/\sqrt{x}$. To freeze this mode, we need to require that the relevant light-cone energy scale satisfies
\begin{equation}\label{eq:finaleregime}
	E_{LC} \ll \frac{R}{\sqrt{x}}\ .
\end{equation}
This is the most stringent of all energy bounds so far, equations~(\ref{eq:Eregime}) and~(\ref{eq:ERN}) -- replacing both, {\em i.e.} $\delta \xi^3$ is the lightest of all off-diagonal bosonic modes. Assuming that the bound~(\ref{eq:finaleregime}) is satisfied by the qubit dynamics of interest and is needed for a notion of emergent spacetime geometry, the $\delta \xi^3$ mode can then also be eventually frozen in the vacuum yielding
\begin{equation}
	\left<\delta \xi^{3\dagger} \delta \xi^3\right> = \frac{\sqrt{r}}{2} \sqrt{\frac{x_1 x_2/r^2}{x_1/r+x_2/r}}
\end{equation}
to leading order in $r/x$. Note that this vev {\em grows} as $\sqrt{x/r}$ and arises from fast modes that straddle the sphere and probes coupling to slower ones straddling the two probes. However, it can be easily checked that the perturbative expansion remains valid. 

Hence, we have identified an energy scale~(\ref{eq:finaleregime}) within the Matrix decoupling regime~(\ref{eq:matrixregime}) where shell and probe qubit effective dynamics arises from integrating out {\em all} off-diagonal fluctuations\footnote{We will see later that, in this energy regime, all off-diagonal fermionic modes are also heavy.}. Otherwise, the strings stretching between the probes become light and a notion of non-commutative geometry sets in.

Let us collect all conditions we have used so far in one place
\begin{eqnarray}
	x \gg r\,N\ \ &&\mbox{Probes far away and in weak gravity} \nonumber \\
	x \gg 1\ \ \ ,\ \ \ \varepsilon \gg 1\ \ &&\mbox{All distances are super-Planckian} \nonumber \\
	x \gg r \gg 1\ \ &&\mbox{Expansion of quartic potential well-defined} \nonumber \\
	\varepsilon \ll x\ \ \ &&\mbox{Probes close enough for detecting tidal effects} \nonumber \\
	E_{LC} \ll \frac{R}{\sqrt{x}}\ \ \ &&\mbox{Energy scale is such that all off-diagonal modes are heavy}\label{eq:mainreg}
\end{eqnarray}
Note that these conditions lie within the Matrix theory decoupling regime~(\ref{eq:matrixregime}). A non-trivial feature of the energy scale $R/\sqrt{x}$ is that it is $N$ and $r$ independent. It may also seem counter-intuitive that the effect of the background shell is to render the strings joining the two probes lighter with increasing distance $x$. However, we must remember that this energy scale should be compared to the gravitational interaction energy scale -- to gauge whether there is a notion of emergent geometry or not. This gravitational interaction is bound to scale as $1/x$, becoming smaller faster than $1/\sqrt{x}$. This implies that the $1/\sqrt{x}$ is consistent with the view that the emergence spacetime and traditional gravity becomes more robust at larger distances where gravity weakens.

Postponing a more detailed discussion of the implications of~(\ref{eq:mainreg}) to the Discussion section,  we now come back to the original task of computing entanglement entropy and hence move onto the fermionic sector. We will see that all off-diagonal fermionic fluctuations are heavy in the regime of interest; we will then integrate them out, and only after that freeze the heavy bosonic modes in their vacua as determined in this section. We will end up with the effective Hamiltonian for the qubits on the membrane and on the probes only.

\subsection{The qubit sector}

From~(\ref{eq:fermions}), our qubit sector corresponds to the following degrees of freedom:
\begin{equation}
	\begin{array}{lcl}
		\delta \psi_{k}^\pm, \overline{\delta \psi}_{k}^\pm & \mbox{  } & \mbox{Off diagonal modes between shell and probes} \\
		\psi^\pm, \overline{\psi}^\pm & \mbox{  } & \mbox{Off diagonal modes between the two probes} \\
		\psi^\pm_k & \mbox{  } & \mbox{Modes associated with the probes} \\
		\Psi^\pm_{pq} & \mbox{  } & \mbox{Modes associated with the shell}
	\end{array}\ .
\end{equation}

The shell-probes qubit entanglement is packaged in the last term of~(\ref{eq:bfssL}). Using the decomposition given in~(\ref{eq:fermions}) and~(\ref{eq:bosperturb}), we write the resulting Hamiltonian in three parts
\begin{equation}\label{eq:Htot}
	H=H_1+H_2+H_3
\end{equation}
with
\begin{eqnarray}\label{eq:H1}
	H_1 &=& R\,r \left(\frac{x^i_{1}}{r}-\frac{x^i_{2}}{r}\right) \left(\psi^{\dagger }\cdot
	   \sigma^i\cdot \psi-\overline{{\psi}}^{\dagger }\cdot \sigma^i\cdot \overline{\psi}\right) \nonumber \\
	   &+& R\, \left(
	(\psi^\dagger_1-\psi^\dagger_2)\cdot\sigma^i\cdot\overline{\psi}\, \delta \xi^i
		   - \psi^\dagger\cdot\sigma^i\cdot(\psi_1- \psi_2)\, \delta \xi^i +\mbox{c.c.}
	\right)
\end{eqnarray}
coupling the qubits on the two probes; and
\begin{eqnarray}
	H_2&=&R\,r \left(\overline{\delta \psi_k}\cdot \sigma^{i\,T}\cdot
	   \left(L_i-\frac{x^i_{k}}{r}\right)\cdot 
	   \overline{\delta \psi_k}^{\dagger } + {\delta \psi_k}^{\dagger }\cdot \sigma^i\cdot \left(L_i-\frac{x^i_{k}}{r}\right)\cdot
	   \delta \psi_k \right) \nonumber \\
   &-& 2\,R\, \left(
   	\overline{\delta \psi}_2 \cdot \sigma^{i\,T} \overline{\delta \psi}_1^\dagger \delta \xi^i + {\delta \psi}_2 \cdot \sigma^i {\delta \psi}_1^\dagger \delta \xi^i + \mbox{c.c.}
   \right) \nonumber \\
	&+& R\,\left[-{\delta x}^{i\dagger}_{k}\cdot \left(\Psi_k^{\dagger}\cdot \sigma^i\cdot
	   \delta \psi_k
	   + \Psi_k\cdot \sigma^{i\,T}\cdot
	   	   \overline{\delta \psi_k}^{\dagger } \right) \right.
	    \nonumber \\
	   &+& {\delta x}^{i\dagger}_{1}\cdot \left(
	   \overline{\psi}\cdot \sigma^{i\,T}\cdot
	   \overline{\delta \psi_2}^{\dagger } 
   		+\psi^{\dagger }\cdot \sigma^i\cdot \delta \psi_2 \right)
	   + {\delta x}^{i\dagger}_{2}\cdot \left(\psi\cdot \sigma^{i\,T}\cdot
	   	   \overline{\delta \psi_1}^{\dagger }
	   +\overline{\psi}^{\dagger }\cdot \sigma^i\cdot \delta \psi_1\right)\nonumber \\
	   &+& \left. \mbox{c.c.} \right]\label{eq:h2full}
\end{eqnarray}
where we define
\begin{equation}
\Psi_k \equiv \Psi -\psi_k\ .
\end{equation}
Note that $H_2$ couples the off-diagonal bosonic modes to the probe and shell qubits. 
And finally, we have
\begin{equation}\label{eq:H3}
	H_3=R\,r\,\mbox{Tr}\left[\Psi ^{\dagger }\cdot \sigma^i\cdot L_i\cdot \Psi \right]
	-R\,r\, \mbox{Tr}\left[\Psi ^{\dagger }\cdot \sigma^i\cdot \Psi \cdot L_i\right]
\end{equation}
which couples the shell qubits to each other. 

Writing $H_1$ in $SU(2)$ component form, 
one gets
\begin{eqnarray}
	H_1&=&R\,r\,\frac{\varepsilon}{r}\left(\psi^{+\dagger}\psi^+-\psi^{-\dagger}\psi^-
-\overline{\psi}^{+\dagger}\overline{\psi}^++\overline{\psi}^{-\dagger}\overline{\psi}^- \right) \nonumber \\
	&+& \sqrt{2}\, R\, \left[
	(\Delta^{-\,\dagger}\overline{\psi}^+-\psi^{-\,\dagger}\Delta^+) \delta \xi^+
	+(\Delta^{+\,\dagger}\overline{\psi}^--\psi^{+\,\dagger}\Delta^-) \delta \xi^-  \right. \nonumber \\
	&-&\left. (\Delta^{-\dagger}\overline{\psi}^--\psi^{-\dagger}\Delta^- - \Delta^{+\dagger}\overline{\psi}^+ + \psi^{+\dagger}\Delta^+)\, \delta \xi^3 + \mbox{c.c.}
	\right]\ , \label{eq:H1expanded}
\end{eqnarray}
where $\varepsilon= x_1-x_2 > 0$ and we define
\begin{equation}
	\Delta^\pm \equiv \frac{1}{\sqrt{2}}\left(\psi^\pm_1-\psi^\pm_2\right)\ .
\end{equation}
The off-diagonal qubits $\psi^\pm$ and $\overline{\psi}^\pm$ are heavy, with energy scaling as $R\,\varepsilon$. We will hence integrate out these modes. However, there is $\psi^\pm$ and $\overline{\psi}^\pm$ dependence in $H_2$ and thus we need to look at $H_2$ before we handle the $\psi^\pm$ and $\overline{\psi}^\pm$ modes.

$H_2$ involves the off-diagonal modes between probes and shell. The modes, $\delta \psi_k$ and $\overline{\delta \psi}_k$, are also heavy -- with frequency scaling as $R\,x_k$ -- and can be integrated out. Hence, we need to diagonalize $H_2$ in these modes. Before we proceed, we can make a useful simplification: the terms on the second line of~(\ref{eq:h2full}) mixing $\delta \psi_1$ and $\delta \psi_2$ are sub-leading to the terms on the first line; they result in shifting the energy of the off-diagonal modes by an amount $r/x$ smaller, which then adds a sub-leading correction to the qubit-qubit entanglement in the final effective Hamiltonian. Hence, we drop these terms from the outset to avoid unnecessary clutter. This allows us to focus on solving a simpler eigenvalue problem
\begin{equation}
	\sigma^i\cdot \left(L_i-\frac{x^i_{k}}{r}\right)\cdot \delta \psi_k =\lambda\,\delta \psi_k \ \ \ , \ \ \ 
	\sigma^{i\,T}\cdot \left(L_i-\frac{x^i_{k}}{r}\right)\cdot \overline{\delta
	   \psi }^{\dagger }_k=\overline{\lambda }\, \overline{\delta \psi }^{\dagger }_k\ .
\end{equation}
Writing $\delta \psi_k$ and $\overline{\delta \psi}_k$ explicitly in spin-$J$ and $SU(2)$ representation components
\begin{equation}
	\delta \psi_{k\,m} = \left(
	\begin{array}{c}
	 \delta \psi^+_{k\,m} \\
	 \delta \psi^-_{k\,m} \\
	\end{array}
	\right)\ ,
\end{equation}
we end up with 
\begin{equation}
	\sqrt{j (j+1)-m (m+1)}\, \delta \psi^-_{k\,m}-
	   \left(\lambda -m+\frac{x_k}{r}-1\right)\,\delta \psi^+_{k\,m+1} =0
\end{equation}
and
\begin{equation}
	\sqrt{j (j+1)-m (m+1)}\, \delta \psi^+_{k\,m+1}-
	   \left(\lambda +m-\frac{x_k}{r}\right)\,\delta \psi^-_{k\,m} = 0
\end{equation}
where we also write $x^3_{k}\equiv x_k$ since the probes are arranged along the $3$ axis. The eigenvalues $\lambda$ can easily be found as
\begin{equation}
	{\lambda}^{(1)}_{m}(x_k)=\frac{1}{2} \left(1-\sqrt{(2
	   j+1)^2+4 \frac{x_k}{r}  \left(-2 m+\frac{x_k}{r} -1\right)}\right)
\end{equation}
and
\begin{equation}
	{\lambda}^{(2)}_{m}(x_k)=\frac{1}{2} \left(1+\sqrt{(2
	   j+1)^2+4 \frac{x_k}{r}  \left(-2 m+\frac{x_k}{r} -1\right)}\right)\ .
\end{equation}
Similarly, for the $\overline{\delta \psi}_k$ degrees of freedom, we get 
\begin{equation}
	{\bar{\lambda}^{(1)}_{m}}(x_k)=-{\lambda}^{(2)}_{m}(x_k)\ \ \ ,\ \ \ {\bar{\lambda}^{(2)}_{m}}(x_k)=-{\lambda}^{(1)}_{m}(x_k)\ .
\end{equation}
For large $x/r\gg 1$, these eigenvalues scale as
\begin{equation}\label{eq:largeevalues}
	\lambda \sim \frac{x}{r}\ ,
\end{equation}
confirming that they are heavy in the regime of interest~(\ref{eq:finaleregime}) -- as already analyzed in~(\ref{eq:Hscaling}).
Using the corresponding eigenvectors, we can write the Hamiltonian in diagonal form in terms of the new variables $\delta \eta_{k\,m}$, $\delta \chi_{k\,m}$, $\overline{\delta \eta}_{k\,m}$, and $\overline{\delta \chi}_{k\,m}$ using
\begin{equation}
	\delta \psi_{k\,m}=\left(
	\begin{array}{c}
	  k^+_{1\,m+1}(x_k) \, \delta \eta_{k\,m+1} + k^+_{2\,m+1}(x_k) \, \delta \chi_{k\,m+1} \\
	  k^-_{1\,m}(x_k) \, \delta \eta_{k\,m} +  k^-_{2\,m}(x_k) \, \delta \chi_{k\,m}  \\
	\end{array}
	\right)
\end{equation}
and
\begin{equation}
	\overline{\delta \psi}^{\dagger }_{k\,m}=\left(
	\begin{array}{c}
	   \overline{k}^+_{1\,m}(x_k) \, \overline{\delta \eta}^{\dagger}_{k\,m} 
	   + \overline{k}^+_{2\,m}(x_k) \, \overline{{\delta \chi}}_{k\,m}^{\dagger } \\
	   \overline{k}^-_{1\,m-1}(x_k) \, \overline{\delta \eta}^{\dagger}_{k\,m-1} 
	   + \overline{k}^-_{2\,m-1}(x_k) \, \overline{{\delta\chi}}^{\dagger}_{k\,m-1} \\
	\end{array}
	\right)\ ,
\end{equation}
where we define the constants
\begin{equation}
	k^+_{i\,m}(x_k)=\frac{\sqrt{(j-m)
	   (j+m+1)}}{\sqrt{(j-m) (j+m+1)+\left({\lambda}^i_m(x_k)-m+(x_k/r)
	   -1\right){}^2}}\ ;
\end{equation}
\begin{equation}
	k^-_{i\,m}(x_k)=\frac{{\lambda}^i_m(x_k )-m+(x_k/r)
	   -1}{\sqrt{(j-m) (j+m+1)+\left({\lambda}^i_m(x_k)-m+(x_k/r)
	   -1\right){}^2}}\ ;
\end{equation}
\begin{equation}
	\overline{k}^+_{i\,m}(x_k)=\frac{{\bar{\lambda }^i_{m}}(x_k)+m-(x_k/r) +1}{\sqrt{(j-m) (j+m+1)+\left({\bar{\lambda}^i_{im}}(x_k)+m-(x_k/r)
	   +1\right){}^2}}\ ;
\end{equation}
\begin{equation}
	\overline{k}^-_{i\,m}(x_k)=\frac{\sqrt{(j-m)
	   (j+m+1)}}{\sqrt{(j-m) (j+m+1)+\left({\bar{\lambda}^i_{m}}(x_k)+m-(x_k/r)
	   +1\right){}^2}}\ .
\end{equation}
For large $x/r\gg 1$, these constants scale as
\begin{equation}
	k^+_1\sim 1\ \ ,\ \ k^-_2 \sim 1\ \ , \ \ \overline{k}^+_1\sim -1\ \ ,\ \ \overline{k}^-_2 \sim 1
\end{equation}
while other $k$'s scale as $r/x$.
Hence, $H_2$ takes the diagonal form
\begin{eqnarray}\label{eq:H2J}
	H_2 &=& R\,r \left({\bar{\lambda}^{(1)}_{m}}(x_k) \overline{\delta \eta}_{k\,m} \overline{\delta \eta}_{k\,m}^{\dagger }
	   +\bar{\lambda}^{(2)}_{m}(x_k) \overline{\delta \chi}_{k\,m} \overline{\delta \chi}_{k\,m}^{\dagger }\right. \nonumber \\
	   && \left. 
	   +\lambda^{(1)}_{m}(x_k) \delta \eta_{k\,m}^{\dagger } \delta \eta_{k\,m}
	   +{\lambda^{(2)}_{m}}(x_k) {\delta \chi_{k\,m}}^{\dagger} \delta \chi_{k\,m}\right) \nonumber \\
	   &+& R\,\left(
   	\delta \eta_{k\,m}^{\dagger } {J\eta}_{k\,m}
   	+\delta \chi_{k\,m}^{\dagger } J\chi_{k\,m} 
   	+{\overline{J\eta}}_{k\,m}  \delta {\overline{\eta}}_{k\,m}^{\dagger }
   	+{\overline{J\chi}}_{k\,m}  \delta {\overline{\chi}}_{k\,m}^{\dagger }
   	+ \mbox{c.c.}
	   \right)
\end{eqnarray}
where we defined the currents ${J\eta}_{k\,m}$, $J\chi_{k\,m}$, ${\overline{J\eta}}_{k\,m}$, and ${\overline{J\chi}}_{k\,m}$. The explicit forms of these currents in terms of $\delta x$ and qubit degrees of freedom on the probes and shell are given in Appendix B. Note in particular that no $\delta \xi^i$ mode appear in these currents.

We can now proceed with integrating out the $\delta \eta_{k\,m}$, $\delta \chi_{k\,m}$, $\overline{\delta \eta}_{k\,m}$, and $\overline{\delta \chi}_{k\,m}$ in $H_2$. For a Hamiltonian in Grassmanian variables $f$ of the form
\begin{equation}
	H=\lambda {f}^\dagger f + J^\dagger f + f^\dagger J
\end{equation}
the resulting partition function takes the form
\begin{equation}\label{eq:effectiveH}
\ln Z_F=-\lim _{\lambda \to \infty }\int ds\,dt\, J(s) J^{\dagger }(t) \theta (t-s)
   e^{i \lambda  (t-s)} 
\end{equation}
with the resulting effective Hamiltonian becoming
\begin{equation}
	H^{{eff}}=i\, \ln Z_F\ .
\end{equation}
In general, this effective Hamiltonian would be non-local in time as we can see from~(\ref{eq:effectiveH}). However, for large $x/r\gg 1$, the $\lambda$ is large as seen from~(\ref{eq:largeevalues}) so that we can write the effective local Fermi-like interaction
\begin{equation}\label{eq:logZf}
	\ln Z_F=i \int dt\,\frac{ J^{\dagger }(t)\cdot J(t)}{\lambda }\ .
\end{equation}
In the case at hand, the effective Hamiltonian~(\ref{eq:H2J}) becomes
\begin{eqnarray}\label{eq:H2effJs}
	H_2^{eff}&=& \frac{R}{r}\left(\frac{J\eta^{\dagger }_{k\,m} J\eta_{k\,m}}{\lambda^{(1)}_{m}(x_k)}
	   +\frac{J\chi{\dagger }_{k\,m} J\chi_{k\,m}}{\lambda^{(2)}_{m}(x_k)} +\frac{\overline{J\eta }_{k\,m} \overline{{J\eta}}^{\dagger }_{k\,m}}{\overline{\lambda}^{(1)}_{m}(x_k)} 
   +\frac{\overline{J\chi }_{k\,m} \overline{J\chi }^{\dagger}_{k\,m} }{\overline{\lambda}^{(2)}_{m}(x_k)}\right)\ ,
\end{eqnarray}
where the currents $J$ are given explicitly in Appendix B. The general form of the currents is
\begin{equation}
	J\sim \delta X \psi
\end{equation}
where $\psi$ represents a qubit mode on the probes or the shell.  Hence the effective Hamiltonian $H_2^{eff}$ couples qubits on the probes and the shell, taking the form
\begin{equation}
	H_2^{eff} \sim \frac{R}{r} \frac{\delta X \delta X}{x/r}\, \psi \psi \sim \frac{R}{r^2} \frac{\psi\psi}{(x/r)^2}\ .
\end{equation}
Let us next look at the last piece of~(\ref{eq:Htot}): $H_3$, given by~(\ref{eq:H3}), involves a dense network of couplings between the $8\,N^2$ shell qubits. To better organize these qubits, we decompose first in matrix spherical harmonics~\cite{hoppe,Dasgupta:2002hx}
\begin{equation}
	\Psi^\pm=\Psi^\pm_{j\,m} Y_{j\,m}\ .
\end{equation}
This allows us to diagonalize $H_3$ by solving the eigenvalue problem given by
\begin{equation}
	\sqrt{(j-m+1) (j+m)} \Psi
	   ^+_{j\,m-1}=(\lambda +m) \Psi
	   ^-_{j\,m}
\end{equation}
and
\begin{equation}
	\sqrt{(j-m) (j+m+1)} \Psi
	   ^-_{j\,m+1}=(\lambda -m) \Psi
	   ^+_{j\,m}
\end{equation}
The solution writes the shell qubits in terms of new fermionic variables $\eta_{jm}$ and $\chi_{jm}$
\begin{equation}\label{eq:shell1}
	\Psi ^+_{j\,m-1}=\frac{\sqrt{j+m} }{\sqrt{2 j+1}
	   \sqrt{N}}\eta_{j\,m}-\frac{\sqrt{j-m+1} }{\sqrt{2 j+1} \sqrt{N}}\chi_{j\,m}
\end{equation}
and
\begin{equation}\label{eq:shell2}
	\Psi ^-_{j\,m}=\frac{\sqrt{j-m+1} }{\sqrt{2 j+1}
	   \sqrt{N}}\eta_{j\,m}+\frac{\sqrt{j+m} }{\sqrt{2 j+1} \sqrt{N}}\chi_{j\,m}
\end{equation}
with $j=0, \cdots, N-1$ on $\eta_{jm}$ and $j=1,\cdots, N-1$ on the $\chi_{jm}$; and $m=-J,\cdots J$.
$H_3$ then takes the diagonal form
\begin{equation}\label{eq:H3final}
	H_3 = R\,r\, \left(j\, \eta ^{\dagger }_{j\,m} \eta_{j\,m}-(j+1)\, \chi_{j\,m} \chi^\dagger_{j\,m}\right)
\end{equation}
in terms of the new variables. These modes are also heavy in the energy regime of interest~(\ref{eq:finaleregime}). They should then be arranged in their vacuum states, given by
\begin{equation}
	\left<\chi_{j\,m} \chi^\dagger_{j\,m}\right> = 1\ \ \ ,\ \ \ \left<\eta ^{\dagger }_{j\,m} \eta_{j\,m} \right> = 0
\end{equation}
for all $j$ and $m$. However, as discussed in Appendix C, the constraint~(\ref{eq:constrainteq}) implies that the physical state of lowest energy is instead given by
\begin{equation}\label{eq:shellvacuum}
	\left<\chi_{j\,m} \chi^\dagger_{j\,m}\right> =\left<\eta ^{\dagger }_{j\,m} \eta_{j\,m} \right> = 1\ \ \ \mbox{No sum over $j$ and $m$}
\end{equation}
for all $j$ and $m$. $H_3$ then yields a zero point energy of $-R\,r\,8\,N^2$. This would cancel against the zero point energies of the bosonic fluctuations of the shell provided we stabilized the shell appropriately. This constant energy shift does not contribute to the entanglement of qubits. 

Let us put together what we have so far. $H_1$ from equation~(\ref{eq:H1expanded}) must be added to the effective Hamiltonian $H_2^{eff}$ given by~(\ref{eq:H2effJs}) and $H_3$ from~(\ref{eq:H3final}). The former is given in terms of the currents listed in Appendix B involving the shell qubit variables as matrix components in the {\em adjoint} representation of $U(N)$, {\em i.e.} $\Psi\rightarrow \Psi_{pq}$ with $p,q=1,\cdots N$. In $H_3$ however, the diagonal form naturally lends itself to matrix spherical harmonics decomposition. Hence, it is convenient to write  $H_2^{eff}$ in terms of $\eta_{jm}$ and $\chi_{jm}$ as well using
\begin{equation}
\left(\Psi _{p\,q}\right){}^\pm=\left(\Psi _{j\,m}\right){}^\pm
   Y_{{j\,m},{p\,q}}=\sqrt{2 j+1} \sqrt{2 J+1} (-1)^{J-p} \left(\Psi
   _{j\,m}\right){}^\pm \left(
\begin{array}{ccc}
 J & j & J \\
 -p & m & q \\
\end{array}
\right)
\end{equation}
with the help of $3j$ symbols; and then by using~(\ref{eq:shell1}) and~(\ref{eq:shell2}) to write the $\Psi$ dependence in the currents $J$ in terms of $\eta_{jm}$ and $\chi_{jm}$.

Before writing the final full effective Hamiltonian, there is one more issue we delayed tackling and that we now need to deal with: The off-diagonal modes $\psi^\pm$ and $\overline{\psi}^\pm$ from $H_1$ are heavy and can be integrated out in the regime of interest~(\ref{eq:finaleregime}); and this steps involves terms in $H_{2}^{eff}$ depending on $\psi^\pm$ and $\overline{\psi}^\pm$. Focusing on the $\psi^\pm$- and $\overline{\psi}^\pm$-dependent terms only in the {\em full} Hamiltonian, we find
\begin{eqnarray}
	H^{eff}_{\psi,\overline{\psi}} &=&
	R\,r\,\frac{\varepsilon}{r}\left(\psi^{+\dagger}\psi^+-\psi^{-\dagger}\psi^-
	-\overline{\psi}^{+\dagger}\overline{\psi}^++\overline{\psi}^{-\dagger}\overline{\psi}^- \right) \nonumber \\
	&+& \left[\psi^{-\dagger} J^-+\psi^{+\dagger} J^++\overline{\psi}^{-\dagger} \overline{J}^-+\overline{\psi}^{+\dagger} \overline{J}^+ + \mbox{c.c}\right] \nonumber \\
	&+& (\mathcal{C}_1+\mathcal{C}_2)(\psi^{-\dagger}\psi^- + \overline{\psi}^{+\dagger}\overline{\psi}^+)
	+(\mathcal{C}_1-\mathcal{C}_2)(\psi^{+\dagger}\psi^+ + \overline{\psi}^{-\dagger}\overline{\psi}^-)\label{eq:Hpsi}
\end{eqnarray}
where
\begin{equation}
	J^\pm \equiv - R \Delta \psi^\mp \delta \xi^\mp \mp R \Delta \psi^\pm \delta \xi^3\ \ \ ,\ \ \ 
	\overline{J}^\pm \equiv R \Delta \psi^\mp \delta \xi^{\pm\dagger} \pm R \Delta \psi^\pm \delta \xi^{3\dagger}\ ,
\end{equation}
and
\begin{equation}
	\mathcal{C}_1 \equiv \frac{R\, a_-}{J^2\, r^2 (x/r)} - \frac{R\, \beta}{r^2 (x/r)^2}\ \ \ ,\ \ \ 
	\mathcal{C}_2 \equiv \frac{R\,(\varepsilon/r)\, a_+}{2\,r^2\,(x/r)^2}\ ,
\end{equation}
where $a_\pm$ and $\beta$ are constants independent of $x$, $\varepsilon$, and $J$. The first two lines of~(\ref{eq:Hpsi}) come from $H_1$ given in~(\ref{eq:H1expanded}); the last line comes from $H_2^{eff}$.
To arrive at the latter, we have taken the vev of the Hamiltonian in the bosonic vacuum and used equation~(\ref{eq:vevs1}); we refer the reader to Appendix D for details for determining the constants $a_\pm$ and $\beta$.
We can now integrate out the heavy $\psi^\pm$ and $\overline{\psi}^\pm$ using~(\ref{eq:logZf}) and take vevs for the $\delta x$'s. We are then lead to the final effective Hamiltonian coupling qubits on the shell and the probes: $16$ qubits associated with the two probes, $\psi_k^{\pm}$, and $8\,N^2$ qubits $\eta_{jm}$ and $\chi_{jm}$ associated with the shell\footnote{Remembering that we have suppressed the $SU(4)$ indices on all the spinors.}. To leading order in  $r/x_k \ll 1$, we find
\begin{equation}\label{eq:Hefftot}
	H^{eff} = H_0+V
\end{equation}
with
\begin{equation}
	H_0 = \frac{R\,x\,J^2}{a_-}\, \left(\sqrt{2}\, x^{1/2}+\frac{8}{\varepsilon}\right) \left[\Delta^{+\dagger}\Delta^{+}+\Delta^{-\dagger}\Delta^{-} \right] + R\,r\, \left(j\, \eta ^{\dagger }_{j\,m} \eta_{j\,m}-(j+1)\, \chi_{j\,m} \chi^\dagger_{j\,m}\right)\label{eq:H0final}
\end{equation}
and
\begin{eqnarray}
	V=
	&-&\frac{R}{J} \frac{1}{x^2} \sum_{n,j}\gamma_n \left(\begin{array}{ccc}
   J & j & J \\
   -n & -1 & n+1
   \end{array}\right)\left(\sqrt{j+2}\,\eta^\dagger_{j\,-1} + \sqrt{j-1}\,\chi^\dagger_{j\,-1}\right)\Sigma^+ \nonumber \\
	&-&\frac{R}{J} \frac{1}{x^2}\sum_{n,j}\gamma_n \left(\begin{array}{ccc}
   J & j & J \\
   -n & 1 & n-1
   \end{array}\right)\left(\sqrt{j+2}\,\eta^\dagger_{j\,2} + \sqrt{j-1}\,\chi^\dagger_{j\,2}\right)\Sigma^- \nonumber \\
	&+& \mbox{c.c.}\label{eq:finalveff}
\end{eqnarray}
where we have traded $\psi_k^\pm$ with $k=1,2$ for the new probe qubit variables
\begin{equation}
	\Delta^\pm =\frac{1}{\sqrt{2}}\left(\psi_1^\pm - \psi_2^\pm\right)\ \ \ ,\ \ \ 
	\Sigma^\pm =\frac{1}{\sqrt{2}}\left(\psi_1^\pm + \psi_2^\pm\right)\ .
\end{equation}
We have also defined
\begin{equation}
	\gamma_n \equiv 
   (-1)^{J+n}\sqrt{(J-n)(1+J+n)}\, \alpha_-(J,n)
\end{equation}
with $\alpha_-(J,n)$ shown in Appendix D. Note also the presence of the $3$j symbols arising from the spherical harmonic decomposition. Interestingly, looking at~(\ref{eq:finalveff}), we see that only shell modes with angular momentum $1$ and $2$ participate in entanglement with the probe qubits. This is one of our main results: the effective Hamiltonian of the qubit system.  We are now ready to compute the entanglement between probe qubits and shell qubits.

\subsection{Entanglement entropy}

Given~(\ref{eq:finalveff}), the couplings in this effective Hamiltonian between probe and shell qubits will invariably lead to quantum entanglement between probe and shell, as illustrated in Appendix A. In this case, we first diagonalize the effective Hamiltonian
\begin{eqnarray}
	H&=&-\frac{R\,r^3}{x^4}
\frac{C}{J}	 \left(\underline{\Sigma}^{+\dagger}\underline{\Sigma}^++\underline{\Sigma}^{-\dagger}\underline{\Sigma}^-\right) \nonumber \\
	&+& R\frac{\sqrt{2}J^2\,r x^{3/2}}{a_-} \left(\underline{\Delta}^{+\dagger}\underline{\Delta}^++\underline{\Delta}^{-\dagger}\underline{\Delta}^-\right)
	+R\,r\, \left(j\, \underline{\eta} ^{\dagger }_{j\,m} \underline{\eta}_{j\,m}-(j+1)\, \underline{\chi}_{j\,m} \underline{\chi}^\dagger_{j\,m}\right)\label{eq:Hdiag}
\end{eqnarray}
in terms of the new qubit variables denoted by underlines. $C$ is an $x$ and $J$ independent constant defined in Appendix D. The diagonalization is shown to leading order in $r/x$. For example, we have
\begin{eqnarray}
	{\Sigma}^+ &=& \underline{\Sigma}^+ - \frac{1}{r\,x^2}\sum_{j,n}\frac{\sqrt{j+2}\, \gamma_n}{j} \left(\begin{array}{ccc}
   J & j & J \\
   -n & -1 & n+1
   \end{array}\right) \underline{\eta}_{j\,-1} \nonumber \\
   &+& \frac{1}{r\,x^2}\sum_{j,n}\frac{\sqrt{j-1}\, \gamma_n}{j+1} \left(\begin{array}{ccc}
   J & j & J \\
   -n & -1 & n+1
   \end{array}\right) \underline{\chi}_{j\,-1}\ ,
\end{eqnarray}
\begin{eqnarray}
	{\Sigma}^- &=& \underline{\Sigma}^- - \frac{1}{r\,x^2}\sum_{j,n}\frac{\sqrt{j+2}\, \gamma_n}{j} \left(\begin{array}{ccc}
   J & j & J \\
   -n & 1 & n-1
   \end{array}\right) \underline{\eta}_{j\,-1} \nonumber \\
   &+& \frac{1}{r\,x^2}\sum_{j,n}\frac{\sqrt{j-1}\, \gamma_n}{j+1} \left(\begin{array}{ccc}
   J & j & J \\
   -n & 1 & n-1
   \end{array}\right) \underline{\chi}_{j\,-1}\ .
\end{eqnarray}
and
\begin{equation}
	\Delta^\pm = \underline{\Delta}^\pm\ .
\end{equation}
From~(\ref{eq:Hdiag}), we see that the vacuum configuration satisfies
\begin{equation}
	\left<\underline{\Sigma}^{\pm\dagger}\underline{\Sigma}^\pm\right> = 1\ \ \ ,\ \ \ \left<\underline{\Delta}^{\pm\dagger}\underline{\Delta}^\pm\right> = 0\ .
\end{equation}
This then in turn implies a non-zero vev for the original qubit variables associated with the probes
\begin{equation}
	\left<{\Sigma}^{\pm\dagger}{\Sigma}^\pm\right> = 1 - \frac{C'}{J^3} \frac{r^2}{x^4} \equiv 1 - \mathcal{C} < 1
\end{equation}
to leading order in $r/x$. The correlation is always less than one in the regime of large distance we are working in. To arrive at this expression, we have used the asymptotic forms from Appendix D where $C'$ is defined as an $x$ and $J$ independent constant. We can now write the reduced density matrix for the probe qubits as
\begin{equation}
	\rho' = \frac{1}{Z}e^{-\left( 1 - \frac{C'}{J^3} \frac{r^2}{x^4}\right) \left({\Sigma}^{+\dagger}{\Sigma}^+ + {\Sigma}^{-\dagger}{\Sigma}^-\right)}
\end{equation}
leading to an entanglement entropy of 
\begin{equation}\label{eq:finalS}
	S = -\ln\left(1-\mathcal{C}\right)-\mathcal{C}\,\ln \left(\frac{\mathcal{C}}{1-\mathcal{C}}\right) > 0
\end{equation}
This is our final result for the entanglement entropy between qubits on the probes and qubits on the shell: the entanglement between supergravity modes on the probes and on the spherical membrane. Notice that we are left with no $\varepsilon$ dependence: at the order we have carried out the computation, the $\varepsilon$ dependence drops out. This expression decreases for larger $x$, and increases with radius $\mathcal{R}$ or the mass $M$ of the source -- with everything else being fixed. Beyond these observations, the structure of the entropy expression is rather opaque. It is worthwhile emphasizing that the energy scale associated with the final effective Hamiltonian~(\ref{eq:Hdiag}) is much larger than~(\ref{eq:finaleregime}): hence, all the qubits ended up frozen in the lowest energy configurations which leads to the entranglement entropy~(\ref{eq:finalS}).

\section{Discussion and outlook}\label{sec:conclusion}
\label{sub:conclusion}

We started this work by affirming that the phenomenon of spacetime emerging from Matrix degrees of freedom involves an energy regime in the theory where all off-diagonal matrix modes are heavy and frozen. This was inspired by numerous computations and suggestions in the literature. We will refer to this as the {\em criterion of emergent Matrix geometry}. The main results of this work can then be summarized as follows:
\begin{itemize}
	\item We have identified a very interesting and novel mechanism at play in Matrix theory. A flat direction in the potential of the theory arising at quadratic order is lifted by couplings to fast modes at quartic order. This is akin to the Paul trap phenomenon used in trapping atoms where a fast rotating saddle surface generated by electromagnetic fields effectively leads to a stable saddle point. We demonstrated that, to see this phenomenon in Matrix theory, one has to consistently include higher order corrections to the dynamics. Fast modes from strings stretched between the shell and the probes couple to slower modes from strings stretching between the probes: the effect is an $x$ dependent mass scale for the strings between the probes which become lighter as we increase $x$ and move the probes further way from the spherical membrane. However, the trend is slower than the fall off of the gravitational potential -- implying that light strings between the probes in fact arise as we get {\em closer} to the source mass, {\em i.e.} as we move to regions of spacetime where the curvature scale increases.
	\item If the criterion of emergent Matrix geometry is satisfied, we have shown a proof of concept computation of Von Neumann entanglement entropy between supergravity modes on a massive spherical membrane and far away super-gravitons. We have shown that integrating out heavy off-diagonal modes generates couplings between qubits on the shell and qubits on the probes; the qubits get furthermore frozen in their vacuum configurations in such a way that non-zero quantum entanglement is found between shell and probes in the vacuum. We have been able to do this computation because of several small dimensionless parameters in the problem that we have used for perturbative expansion: the ratio of the radius of the shell to the probes-shell distance, the ratio of the distance between the two probes to their distance from the shell, the inverse of the units of light-cone momentum  $1/N$, and by taking all distances to be super-Planckian. These were summarized in equation~(\ref{eq:mainreg}). 
\end{itemize}

These conclusions lead us to several challenging issues which we now list:

\begin{itemize}
	\item From the M theory perspective and equation~(\ref{eq:gravregime}), we guessed that the light-cone energy scale that captures gravitating probes must scale as
\begin{equation}\label{eq:elcx}
	E_{LC} \sim \frac{R\,N}{x}\ .
\end{equation}
This implies that we want and need
\begin{equation}\label{eq:xN2}
	\frac{R\,N}{x} \ll \frac{R}{\sqrt{x}} \Rightarrow x\gg N^2
\end{equation}
in order to have a notion of emergent geometry from matrices where off-diagonal matrix modes are heavy {\em and} the probes are gravitating from the dual M theory perspective. This implies that, given the large shell whose horizon is at $x\sim r\,N$, we may expect that the notion of spacetime geometry around it breaks down at a distance $x\sim N^2$ from the shell -- a distance that is far away from the central singularity hidden behind the horizon as long as $r>N$. Indeed, depending on $r$, the breakdown of spacetime geometry can extend outside the horizon: if the spherical membrane radius is made too small, the breakdown of smooth commutative spacetime leaks outside of the horizon. This however is misleading. In arriving to such a conclusion, we use~(\ref{eq:elcx}) which is valid asymptotically far away from the shell; whereas the $x\sim N^2$ distance lies in a regime where the asymptotic condition~(\ref{eq:metricexpansion}) breaks down. Hence, using~(\ref{eq:elcx}) may be inconsistent with the conclusion in~(\ref{eq:xN2}). On the other hand, no matter how equation~(\ref{eq:elcx}) gets modified as we approach the horizon, we expect that the strings connecting the probes do become light far away from regions of Planckian curvature: the reason for this is that, on physical grounds, we should expect any modified expression for $E_{LC}$ to scale as a negative power of $x$ and as a positive power $M$ and/or $N$. Together with the critical energy scale $E_{LC} \sim R/\sqrt{x}$, this necessarily leads to a breakdown of commutative geometry at $x\gg 1$.

\item Intuitively, we may expect that the strings connecting the two probes become light near the horizon, in tune with suggestion in the literature~\cite{Mathur:2005zp}-\cite{Almheiri:2012rt}. To see this, we would need the dispersion relation of the probes near the horizon as opposed to in the regime that is asymptotically far away. This would however lie in an energy regime where our matrix computation may get invalidated. Hence, we cannot explore this possibility within the current work. Neverthless, such a picture may have significant implications to the firewall proposal~\cite{Almheiri:2012rt}, even though to make the connection, one needs to, in addition, understand how to switch perspective to the in-falling observer viewpoint in Matrix theory language.   
	
	\item By the time we got to the computation of entanglement entropy, we found no $\varepsilon$ dependence in the final result. This is due to the fact that this dependence arises at a higher order than our computation considered. Hence, to capture a signature of the tidal force effect between the probes, one may need to expand all expressions to higher order in $\varepsilon$ -- which makes the computation significantly more involved. There is however a shortcut to seeing the desired effect with a different setup, as discussed below. At the current level of perturbative expansion, our expression for the entanglement entropy~(\ref{eq:finalS}) is difficult to decipher. There is dependence on distance and mass, but not in a form that makes a map between the entropy and the asymptotic spacetime geometry obvious.
	
\end{itemize}

We have taken the first steps in exploring a fascinating new direction in the program of mapping a dictionary between entanglement entropy, Matrix theory, and emergent geometry. There remains many open directions to pursue to understand the sense in which geometry emerges from D0 brane Matrix degrees of freedom:  

\begin{itemize}
	\item One can consider different arrangements and regimes in which probes are placed far away from a spherical and non-spherical membrane (planar and cylindrical configurations are obvious alternatives), in arbitrary target space dimensions, and one can then compute the corresponding entanglement entropy between supergravity modes on the probes and the source along the line we have outlined in this work. More case studies are bound to clarify a map between local curvature seen by the probes and entanglement entropy -- akin to the holographic area-entropy proposal of~\cite{Ryu:2006bv,Hubeny:2007xt} but now with a more local character and in the case of flat asymptotic background. One particularly interesting scenario is that of two massless probes moving along parallel in-falling geodesics. The convergence rate of such geodesics in known to be related to a local notion of holographic area and a corresponding c-function in a dual picture~\cite{Sahakian:1999bd}. It seems that computation of entanglement entropy in such a setup may be more promising and tractable. We hope to report on this in the near future~\cite{toappear}.
	
	\item It would be very interesting to determine the dynamics of the shell radius, instead of fixing it externally. We have done some preliminary work in trying to understand the nature of the fluctuations in the $N\times N$ block where the spherical membrane is arranged. However, the computation is a challenging one and may necessitate numerical methods. It appears that key to such a mechanism is a back reaction phenomenon of matrix degrees of freedom back upon themselves, along with a Myers-like effective dielectric phenomenon~\cite{Myers:1999ps}. Alternatively,  one may consider similar scenarios in the BMN Matrix theory~\cite{Berenstein:2002jq}, where the spherical fuzzy sphere is classically stable and its radius is fixed by the equations of motion.
	
\end{itemize}

\section{Appendices}

\subsection{Appendix A: Spectral analysis and entropy}

In this appendix, let us illustrate the key to the entanglement mechanism in the vacuum between probes and shell in general terms. Consider a qubit system with Hamiltonian
\begin{equation}\label{eq:genericH}
	H= \sum a_{mn} f_m^\dagger f_n
\end{equation}
for arbitrary $a_{mn}$. Diagonalizing the system through
\begin{equation}\label{eq:overlap}
	F_k = \sum_m c_{km} f_m
\end{equation}
we end up with a Hamiltonian of the form
\begin{equation}
	H = \sum_k \lambda_k F_k^\dagger F_k\ .
\end{equation}
The Fermi vacuum $\left|\Omega\right>$ of the system then has a condensate of fermionic modes for all $\lambda_n < 0$
\begin{equation}
	\left<\Omega \right| F_k^\dagger F_l \left|\Omega\right> = \delta_{kl} \ \ \ \mbox{for all $\lambda_k < 0$}\ .
\end{equation}
This in turn in general implies a non-zero vacuum expectation value for the original fermionic modes
\begin{equation}
	\left<\Omega \right| f_m^\dagger f_n \left|\Omega\right> \neq 0\ .
\end{equation}
for $f_m$ modes that overlap with the excited $F_k$'s as determined from~(\ref{eq:overlap}).
Now imagine that we pick a subset of the $f_m$ modes and ask for the entanglement entropy for these modes with the rest of the system in the vacuum. Because the original system is that of free fermions -- with a Hamiltonian that is quadratic in the $f_m$'s, we can proceed as follows. The reduced density matrix must take the form~\cite{pescheleisler}
\begin{equation}
	\rho' = \frac{1}{Z} e^{-\mathcal{H}} = \frac{1}{Z} e^{-h_{mn} f^\dagger_m f_n}
\end{equation}
where $Z$ is the normalization constant so that $\mbox{Tr}\rho'=1$; and $\mathcal{H}$ is known as the entanglement Hamiltonian. The sum in the exponent includes only the qubits in the subsystem of interest. And the coefficient $h_{mn}$ can be found by computing the relevant correlators from the original Hamiltonian~(\ref{eq:genericH}), that is the $\left<\Omega \right| f_m^\dagger f_n \left|\Omega\right>$'s for $m$ and $n$ in the subsystem of interest. Wick's theorem guarantees that all correlator data is indeed packed in these two-point correlators. It is easier to derive the entanglement Hamiltonian if we diagonalize it so that
\begin{equation}\label{eq:reduced}
	\rho' = \frac{1}{Z} e^{- \sum_k \varepsilon_k F^\dagger_k F_k}
\end{equation}
where the sum is over the subsystem degrees of freedom. Writing
\begin{equation}
	\left<\Omega \right| F_k^\dagger F_l \left|\Omega\right> = c_k \delta_{kl}
\end{equation}
we easily find
\begin{equation}\label{eq:epsilon}
	\varepsilon_k = \ln \frac{1-c_k}{c_k}\ .
\end{equation}
Hence, by computing two point correlators in the original Hamiltonian, we find the $c_k$'s and we can construct the reduced density matrix from~(\ref{eq:reduced}) and~(\ref{eq:epsilon}). And the Von Neumann entropy then takes the standard form 
\begin{equation}\label{eq:vonneumann}
	S = - \mbox{Tr} \rho' \ln \rho' = \sum_{k} \ln \left(1+e^{-\varepsilon_k}\right) + \frac{\varepsilon_k}{e^{\varepsilon_k}+1}\ .
\end{equation}

\subsection{Appendix B: Currents}

In this appendix, we collect the currents coupling to the off-diagonal fermionic modes that get integrated out. Using the notation in the text, we have
\begin{eqnarray}
&&J\eta_{1\,m}=k^-_{1\,m}(x_1) \left(-\delta X^+_{1\,n} {\Psi}^+_{{1\,m\,n}}
    +\delta X_{1\,n} {\Psi}^-_{{1\,m\,n}}+{\overline{\psi}}^+ \delta X^+_{2\,m}-{\overline{\psi}}^- {\delta X}_{2\,m}\right)\nonumber \\
   &&+k^+_{1\,m}(x_1)
   \left(-\delta X^-_{1\,n} {\Psi}^-_{{1\,m+1\,n}}-\delta X_{1\,n} {\Psi}^+_{{1\,m+1\,n}}+\overline{\psi}^- \delta X^-_{2\,m+1}+\overline{\psi}^+
   \delta X_{2\,m+1}\right)
\end{eqnarray}

\begin{eqnarray}
	&&J\eta_{2\,m}=k^-_{1\,m}(x_2) \left(-\delta X^+_{2\,n} {\Psi}^+_{{2\,m\,n}}+\delta X_{2\,n} {\Psi}^-_{{2\,m\,n}}+\psi^+ \delta X^+_{1\,m}-\psi^- {\delta X}_{1\,m}\right)\nonumber \\
   &&+k^+_{1\,m}(x_2) \left(-\delta X^-_{2\,n} {\Psi}^-_{{2\,m+1\,n}}-\delta X_{2\,n} {\Psi}^+_{{2\,m+1\,n}}+\psi^- \delta X^-_{1\,m+1}+\psi^+ {\delta X}_{1\,m+1}\right)
\end{eqnarray}

\begin{eqnarray}
	&&J\chi^\dagger_{1\,m}=k^-_{2\,m}(x_1) \left(\delta X^+_{1\,n} {\Psi}^+_{{1\,m\,n}}+\delta X_{1\,n} {\Psi}^-_{{1\,m\,n}}+\overline{{\psi}}^+ \delta X^+_{2\,m}-\overline{\psi}^- {\delta X}_{2\,m}\right)\nonumber \\
   &&+k^+_{2\,m}(x_1)
	   \left(\delta X^-_{1\,n} {\Psi}^-_{{1\,m+1\,n}}-\delta X_{1\,n} {\Psi}^+_{{1\,m+1\,n}}+\overline{\psi}^- \delta X^-_{2\,m+1}+\overline{\psi}^+
	   \delta X_{2\,m+1}\right)
\end{eqnarray}
\begin{eqnarray}
	&&J\chi_{2\,m}=k^-_{2\,m}(x_2) \left(-\delta X^+_{2\,n} {\Psi}^+_{{2\,m\,n}}+\delta X_{2\,n} {\Psi}^-_{{2\,m\,n}}+\psi^+ \delta X^+_{1\,m}-\psi^- {\delta X}_{1\,m}\right)\nonumber \\
   &&+k^+_{2\,m}(x_2) \left(-\delta X^-_{2\,n} {\Psi}^-_{{2\,m+1\,n}}-\delta X_{2\,n} {\Psi}^+_{{2\,m+1\,n}}+\psi^- \delta X^-_{1\,m+1}+\psi^+ {\delta X}_{1\,m+1}\right)
\end{eqnarray}

\begin{eqnarray}
&&\overline{J\eta }_{1\,m}=\overline{k}^-_{1\,m}(x_1)
   \left(\delta X^{-\dagger}_{1\,n} {\Psi}^+_{{1\,n\,m+1}}+{\delta X^\dagger}_{1\,n}
   \Psi^-_{{1\,n\,m+1}}+\psi^+
   \delta X^{-\dagger}_{2\,m+1}-\psi^- \delta X^\dagger_{2\,m+1}\right)\nonumber \\
   &&+\overline{k}^+_{1\,m}(x_1)
   \left(\delta X^{+\dagger}_{1\,+\,n} {\Psi}^-_{{1\,n\,m}}-\delta X^\dagger_{1\,n} {\Psi}^+_{{1\,n\,m}}+\psi^- \delta X^{+\dagger}_{2\,m}+{\psi}^+ \delta X^\dagger_{2\,m}\right)
\end{eqnarray}
\begin{eqnarray}
	&&\overline{J\eta }_{2\,m}=\overline{k}^-_{1\,m}(x_2)
	   \left(-\delta X^{-\dagger}_{2\,n} {\Psi}^+_{{2\,n\,m+1}}+\delta X^\dagger_{2\,n} {\Psi}^-_{{2\,n\,m+1}}+\overline{\psi}^+
	   \delta X^{-\dagger}_{1\,m+1}-\overline{\psi}^-
	   \delta X^\dagger_{1\,m+1}\right)\nonumber \\
   &&+\overline{k}^+_{1\,m}(x_2)
	   \left(-\delta X^{+\dagger}_{2\,n} {\Psi}^-_{{2\,n\,m}}-\delta X^\dagger_{2\,n} {\Psi}^+_{{2\,n\,m}}+\overline{\psi}^-
	   \delta X^{+\dagger}_{1\,m}+\overline{\psi}^+
	   \delta X^\dagger_{1\,m}\right)
\end{eqnarray}

\begin{eqnarray}
&&\overline{J\chi }_{1\,m}=\overline{k}^-_{2\,m}(x_1)
   \left(\delta X^{-\dagger}_{1\,-\,n} {\Psi}^+_{{1\,n\,m+1}}+\delta X^\dagger_{1\,n}
   \Psi^-_{{1\,n\,m+1}}+\psi^+
   \delta X^{-\dagger}_{2\,-\,m+1}-\psi^- \delta X^\dagger_{2\,m+1}\right)\nonumber \\
   &&+\overline{k}^+_{2\,m}(x_1)
   \left(\delta X^{+\dagger}_{1\,n} {\Psi}^-_{{1\,n\,m}}-\delta X^\dagger_{1\,n} {\Psi}^+_{{1\,n\,m}}+\psi^- \delta X^{+\dagger}_{2\,m}+{\psi}^+ \delta X^\dagger_{2\,m}\right)
\end{eqnarray}
\begin{eqnarray}
	&&\overline{J\chi }_{2\,m}=\overline{k}^-_{2\,m}(x_2)
	   \left(-\delta X^{-\dagger}_{2\,n} {\Psi}^+_{{2\,n\,m+1}}+\delta X^\dagger_{2\,n} {\Psi}^-_{{2\,n\,m+1}}+\overline{\psi}^+
	   \delta X^{-\dagger}_{1\,m+1}-\overline{\psi}^-
	   \delta X^\dagger_{1\,m+1}\right)\nonumber \\
   &&+\overline{k}^+_{2\,m}(x_2)
	   \left(-\delta X^{+\dagger}_{2\,n} {\Psi}^-_{{2\,n\,m}}-\delta X^\dagger_{2\,n} {\Psi}^+_{{2\,n\,m}}+\overline{\psi}^-
	   \delta X^{+\dagger}_{1\,m}+\overline{\psi}^+
	   \delta X^\dagger_{1\,m}\right)
\end{eqnarray}

\subsection{Appendix C: Constraint and physical states}

The constraint given by~(\ref{eq:constrainteq}) translates to statements on the Hilbert space of qubits that picks out the physical states. 
We start with
\begin{equation}
	\delta \Psi^\dagger \cdot \delta \Psi= 0\ .
\end{equation}
Written in our qubit variables, this becomes
\begin{eqnarray}
	&&\eta^\dagger_{jm}\eta_{jm} \left|\mbox{phys}\right> = \chi_{jm} \chi^\dagger_{jm}\left|\mbox{phys}\right> \ \ \ \mbox{no sum over $j$ and $m$}\nonumber \\
	&& \psi_1^{\pm\dagger} \psi^\pm_1\left|\mbox{phys}\right>= \psi^\pm {\psi}^{\pm\dagger}\left|\mbox{phys}\right>\ \ \ , \ \ \ 
	\psi_2^{\pm\dagger} \psi^\pm_2\left|\mbox{phys}\right>=\overline{\psi}^\pm\, \overline{\psi}^{\pm\dagger} \left|\mbox{phys}\right>\ . \label{eq:qubitconstaint}
\end{eqnarray}
with no sum over $j$ and $m$.

\subsection{Appendix D: 3j symbols and asymptotics}

In dealing with matrix spherical harmonics, we invariably encounter Wigner's 3j symbols. To simplify expression, we make use of the following identities. Under even permutations of columns, we have
\begin{equation}
	P_{even}\left(\begin{array}{ccc}j_1 & j_2 & j_3 \\ m_1 & m_2 & m_3 \end{array}\right) = \left(\begin{array}{ccc}j_1 & j_2 & j_3 \\ m_1 & m_2 & m_3 \end{array}\right)\ .
\end{equation}
Under odd permutations of columns, we instead have
\begin{equation}
	P_{odd}\left(\begin{array}{ccc}j_1 & j_2 & j_3 \\ m_1 & m_2 & m_3 \end{array}\right) = (-1)^{j_1+j_2+j_3} \left(\begin{array}{ccc}j_1 & j_2 & j_3 \\ m_1 & m_2 & m_3 \end{array}\right)\ .
\end{equation}
Furthermore, we can also flip lower row signs
\begin{equation}
	\left(\begin{array}{ccc}j_1 & j_2 & j_3 \\ -m_1 & -m_2 & -m_3 \end{array}\right) = (-1)^{j_1+j_2+j_3} \left(\begin{array}{ccc}j_1 & j_2 & j_3 \\ m_1 & m_2 & m_3 \end{array}\right)\ .
\end{equation}
Note also that the $3j$ symbols vanish unless $m_1+m_2+m_3=0$. 

In the text, we also encounter several rather complicated expressions which are only functions of $J$ (or $N$). In the needed regime $J\gg 1$, all these expressions scale as a power of $J$, which we can determine by numerical plots. For this purpose, we define
\begin{equation}
	\alpha_\pm(J,n) \equiv \frac{1}{2}\left(
	\frac{1}{\sqrt{2\,J\,(J+1)-\kappa_{1n}^2-\kappa_{2n}^2}}
	\pm \frac{1}{\sqrt{2\,J\,(J+1)-\kappa_{1n+1}^2-\kappa_{2n+1}^2}}
	\right)
\end{equation}
\begin{equation}
	\alpha(J,n) \equiv \frac{1}{2}\left(
	\frac{n}{\sqrt{2\,J\,(J+1)-\kappa_{1n}^2-\kappa_{2n}^2}}
	- \frac{n+1}{\sqrt{2\,J\,(J+1)-\kappa_{1n+1}^2-\kappa_{2n+1}^2}}
	\right)
\end{equation}
and we find the asymptotic forms for large $J\gg 1$
\begin{equation}\label{eq:as1}
	\sum_{n} \alpha_-(J,n) = \frac{a_-}{J^2}\ \ \ ,\ \ \ 
	\sum_{n} \alpha_+(J,n) = a_+\ \ \ ,\ \ \ 
	\sum_{n} \alpha(J,n) = \beta\ .
\end{equation}
where $a_\pm$ and $\beta$ are $J$ independent numerical constants. To see how to arrive at these asymptotic expressions, we first write
\begin{equation}
	C_1=\sum_{n=-J}^J \alpha_-(J,n)\ \ \ ,\ \ \ C_2=J\, \sum_{n=-J}^J \alpha_+(J,n)\ \ \ ,\ \ \ C_3=J\, \sum_{n=-J}^J \alpha(J,n)
\end{equation}
and numerically plot $\ln |C_{1,2,3}|$ versus $\ln J$. This gives the results quoted in~(\ref{eq:as1}).

\begin{figure}
	\begin{center}
		\includegraphics[width=2.5in]{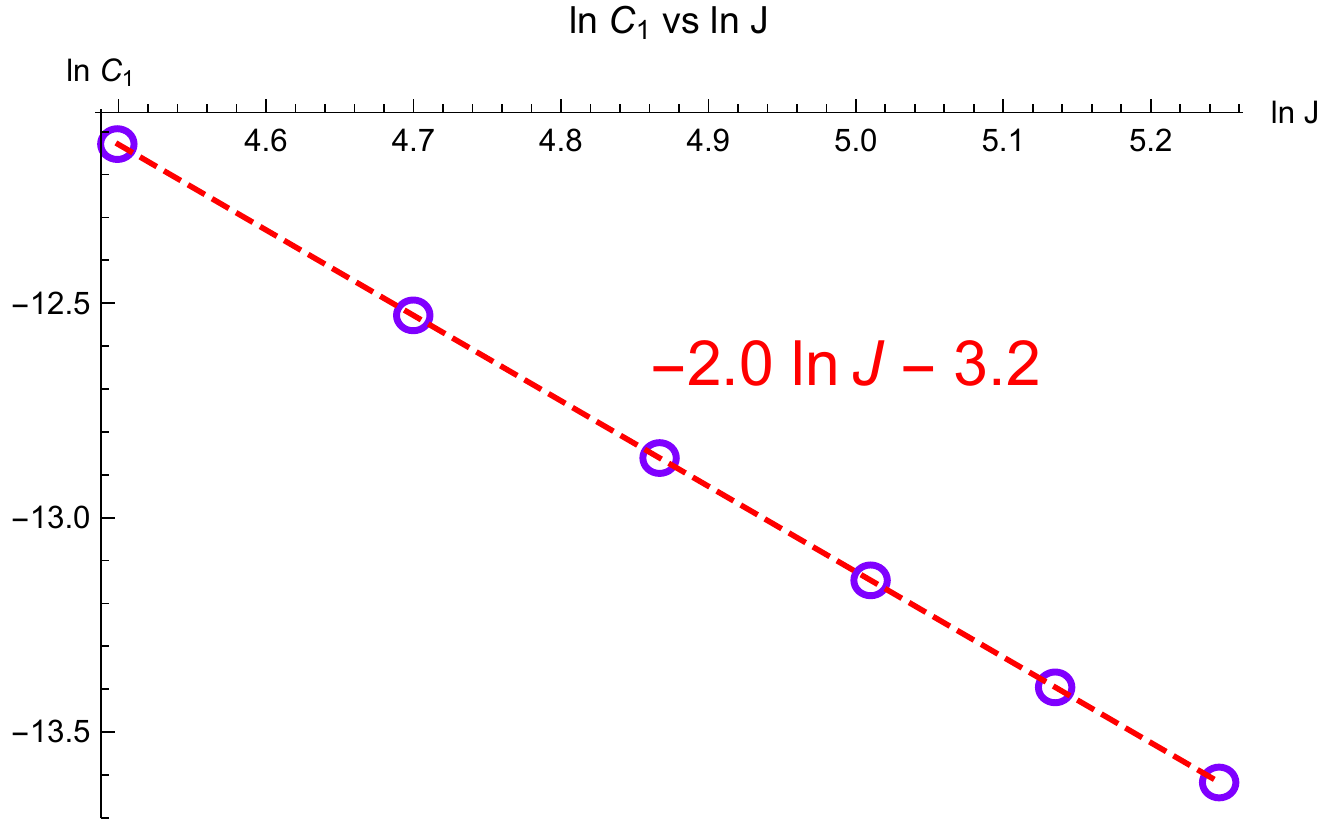}
		\includegraphics[width=2.5in]{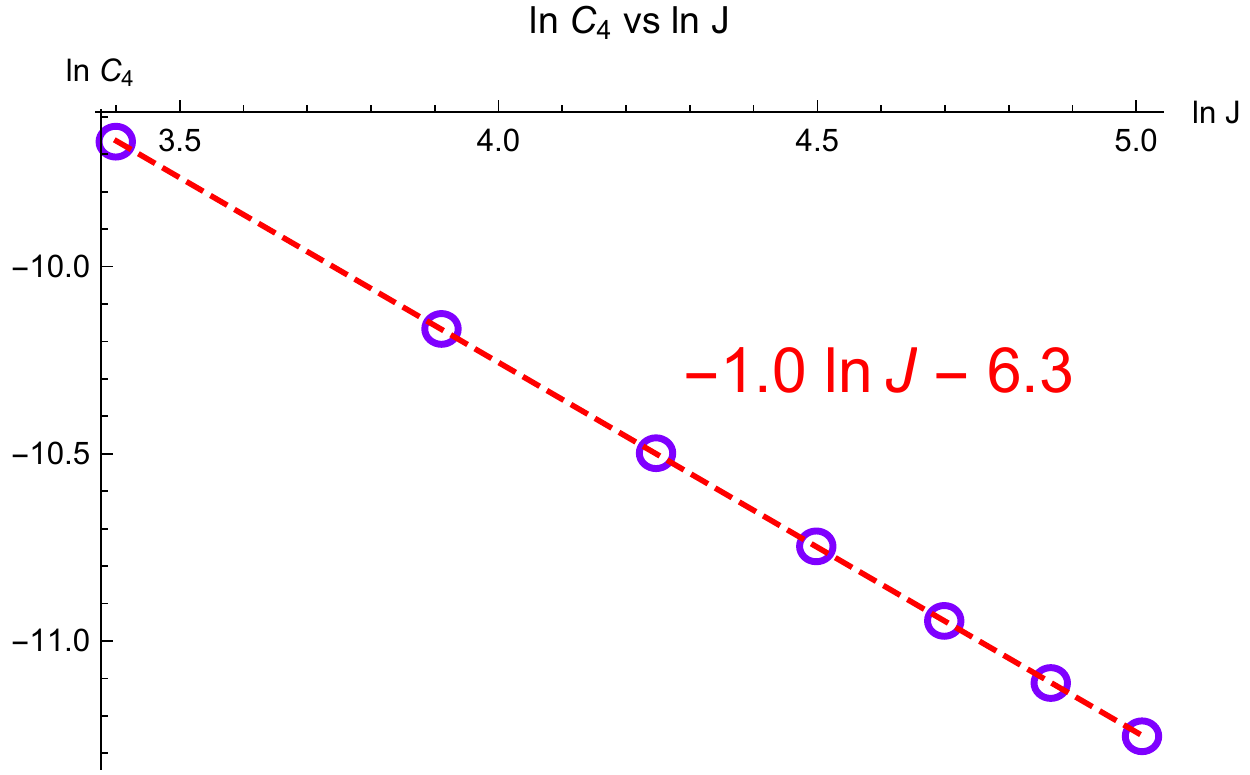}
		\includegraphics[width=2.5in]{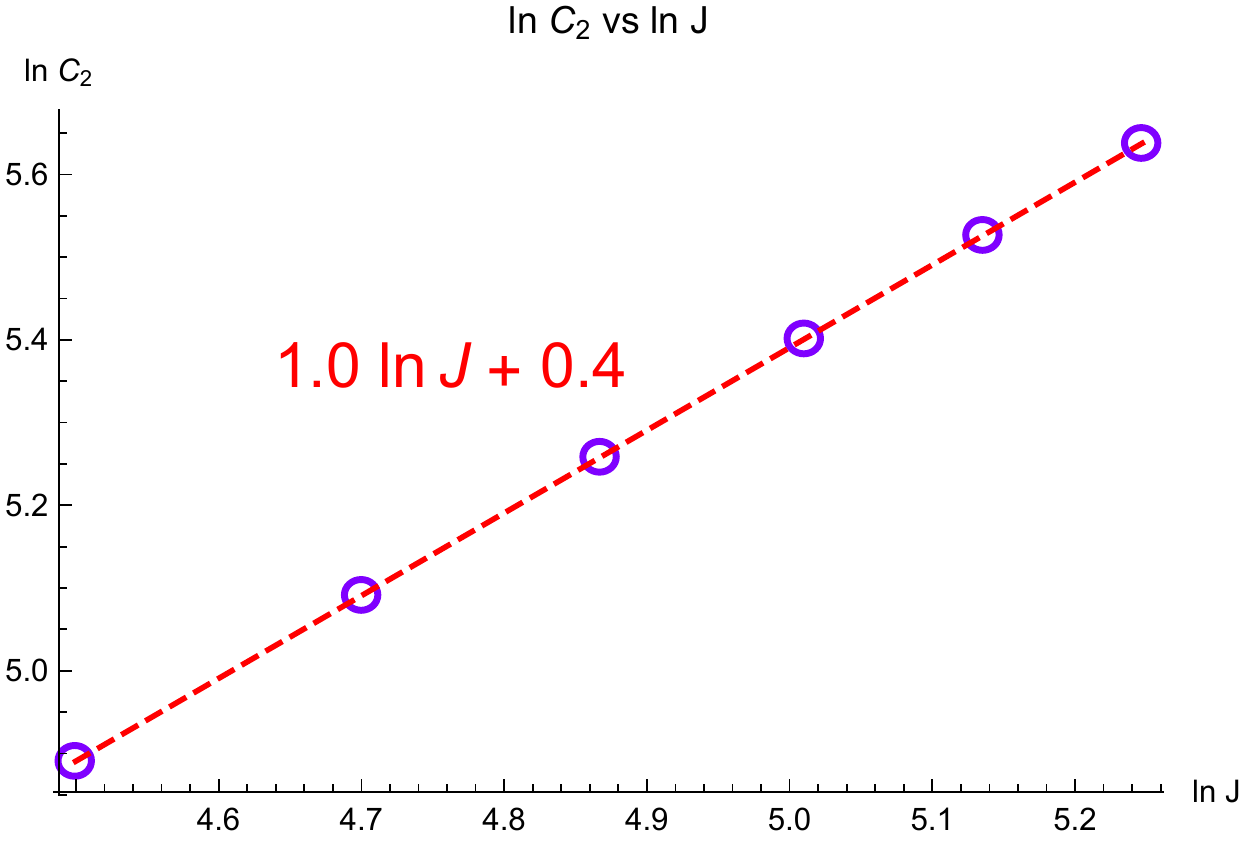}
		\includegraphics[width=2.5in]{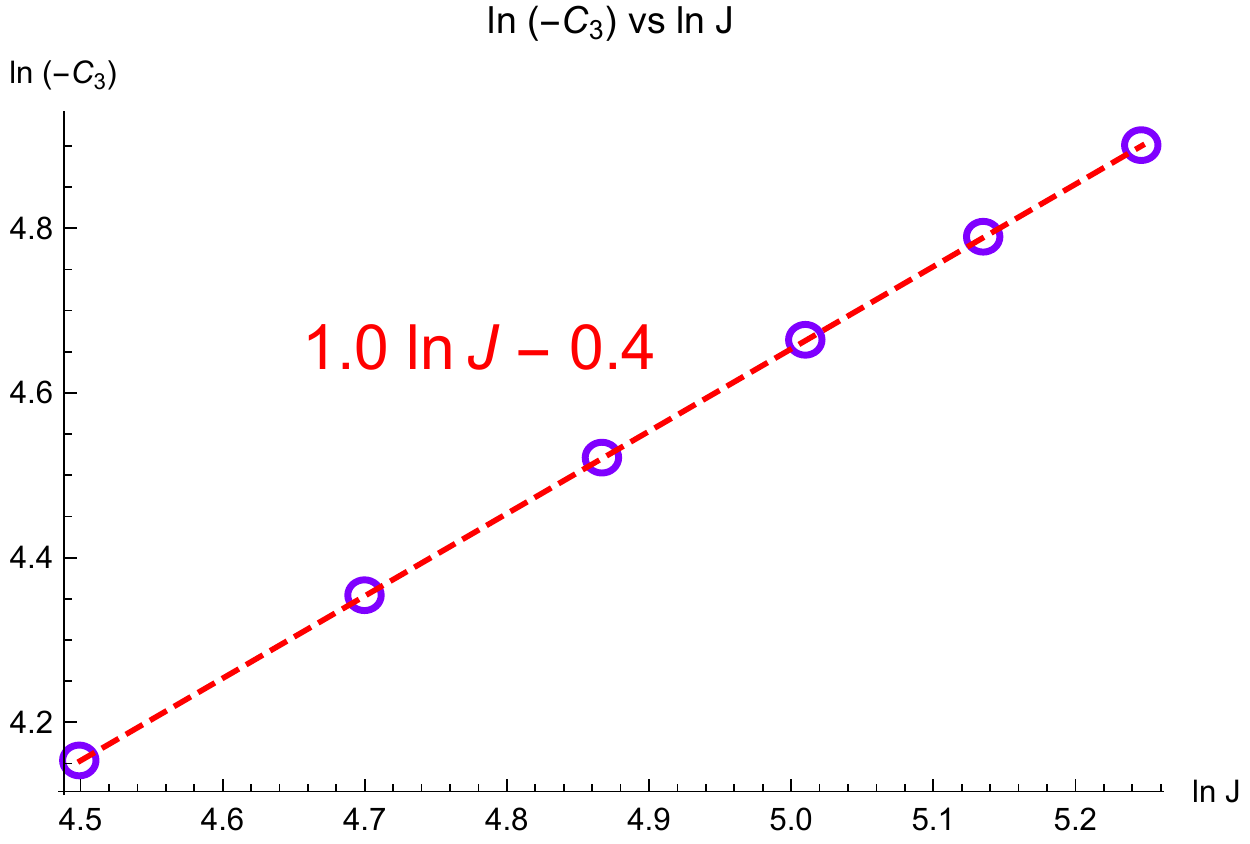}
		\includegraphics[width=2.5in]{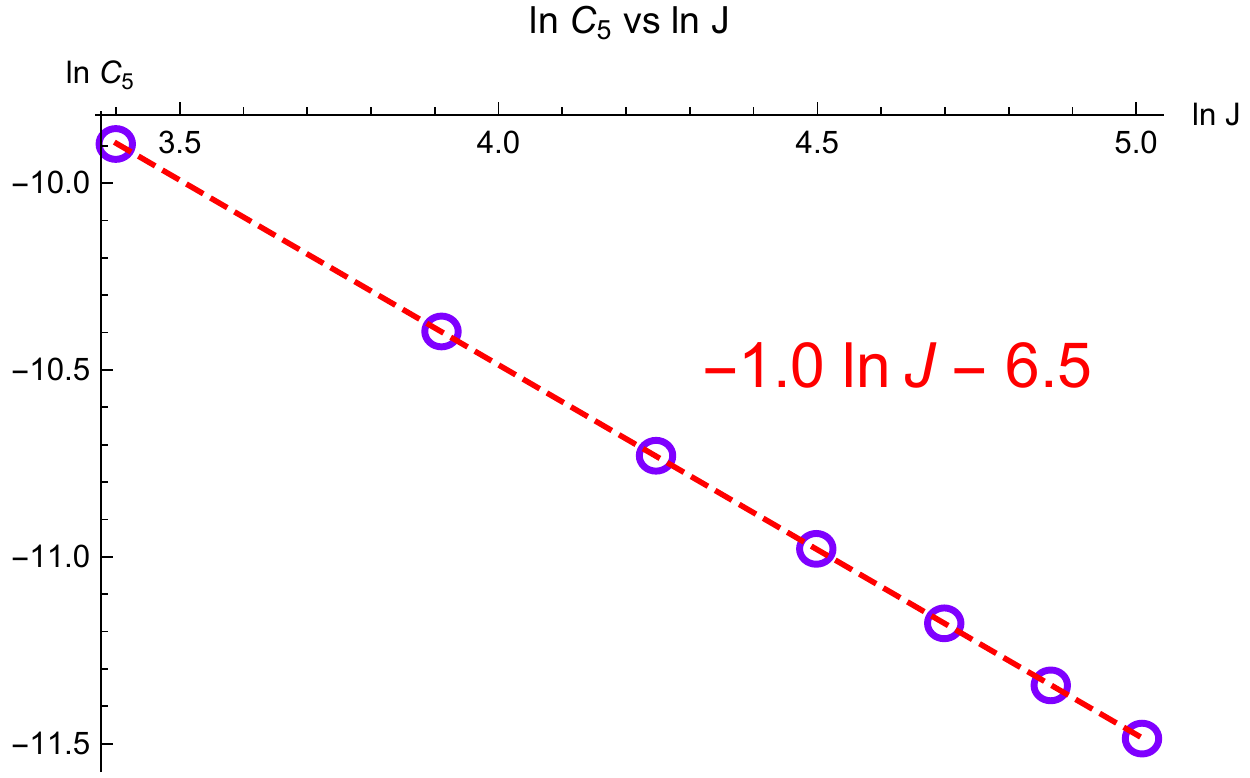}
	\end{center}
	\caption{Numerical plots of $\ln |C_k|$ versus $\ln J$, along with the linear fits. We take modestly large values of $J$ ranging from $30$ to $200$, yet we are able to identify robust power law behavior.}\label{fig:plots}
\end{figure}

Similarly, we define
\begin{equation}
	\gamma_n \equiv (-1)^{J-n} \sqrt{(J-n) (J+n+1)} \, \alpha_-(J,n)
\end{equation}
and write
\begin{equation}
	C_4=-\sum_{n=-J}^J \sum_{j=1}^{2\,J} 
	\frac{4 (2 j+1)}{j (j+1)} 
	{\left| 
\left(
\begin{array}{ccc}
 J & j & J \\
 -n & -1 & n+1 \\
\end{array}
\right)	
	\gamma_n\right| ^2}
\end{equation}
and
\begin{equation}
C_5= \sum_{n,n'=-J}^J\sum_{j=1}^{2\,J} 
\frac{2 (2 j+1) (j (j+1)+2)}{j^2 (j+1)^2} {\gamma_n \gamma_{n'}
\left(
\begin{array}{ccc}
 J & j & J \\
 -n & -1 & n+1 \\
\end{array}
\right)	
\left(
\begin{array}{ccc}
 J & j & J \\
 -n' & -1 & n'+1 \\
\end{array}
\right)	
}
\end{equation}
and find the asymptotic forms
\begin{equation}
	C_4 \simeq \frac{C}{J}\ \ ,\ \ 
	C_5 \simeq \frac{C'}{J}
\end{equation}
through numerical plots, where $C$ and $C'$ are $J$ independent numerical constants. All five of these expressions are shown in Figure~\ref{fig:plots}, with the corresponding fits. We see that we have very robust asymptotic behavior throughout. These results are used throughout the main text.

\newpage
\section{Acknowledgments}

This work was supported by NSF grant number PHY-0968726 and a grant from the Butler Fund.

\providecommand{\href}[2]{#2}\begingroup\raggedright\endgroup


\begin{thebibliography}{10}

\bibitem{Ryu:2006bv}
S.~Ryu and T.~Takayanagi, ``Holographic derivation of entanglement entropy from
  ads/cft,'' {\em Phys. Rev. Lett.} {\bf 96} (2006) 181602,
  \href{http://xxx.lanl.gov/abs/hep-th/0603001}{{\tt hep-th/0603001}}.

\bibitem{Ryu:2006ef}
S.~Ryu and T.~Takayanagi, ``{Aspects of holographic entanglement entropy},''
  {\em JHEP} {\bf 08} (2006) 045,
  \href{http://xxx.lanl.gov/abs/hep-th/0605073}{{\tt hep-th/0605073}}.

\bibitem{Hubeny:2007xt}
V.~E. Hubeny, M.~Rangamani, and T.~Takayanagi, ``A covariant holographic
  entanglement entropy proposal,''
  \href{http://xxx.lanl.gov/abs/arXiv:0705.0016 [hep-th]}{{\tt arXiv:0705.0016
  [hep-th]}}.

\bibitem{Nishioka:2009un}
T.~Nishioka, S.~Ryu, and T.~Takayanagi, ``{Holographic Entanglement Entropy: An
  Overview},'' {\em J. Phys.} {\bf A42} (2009) 504008,
  \href{http://xxx.lanl.gov/abs/0905.0932}{{\tt 0905.0932}}.

\bibitem{VanRaamsdonk:2009ar}
M.~Van~Raamsdonk, ``{Comments on quantum gravity and entanglement},''
  \href{http://xxx.lanl.gov/abs/0907.2939}{{\tt 0907.2939}}.

\bibitem{Maldacena:2013xja}
J.~Maldacena and L.~Susskind, ``{Cool horizons for entangled black holes},''
  \href{http://xxx.lanl.gov/abs/1306.0533}{{\tt 1306.0533}}.

\bibitem{Czech:2014ppa}
B.~Czech and L.~Lamprou, ``{Holographic definition of points and distances},''
  {\em Phys. Rev.} {\bf D90} (2014) 106005,
  \href{http://xxx.lanl.gov/abs/1409.4473}{{\tt 1409.4473}}.

\bibitem{Bao:2015bfa}
N.~Bao, S.~Nezami, H.~Ooguri, B.~Stoica, J.~Sully, and M.~Walter, ``{The
  Holographic Entropy Cone},'' {\em JHEP} {\bf 09} (2015) 130,
  \href{http://xxx.lanl.gov/abs/1505.07839}{{\tt 1505.07839}}.

\bibitem{Banks:1996vh}
T.~Banks, W.~Fischler, S.~H. Shenker, and L.~Susskind, ``M theory as a matrix
  model: A conjecture,'' {\em Phys. Rev.} {\bf D55} (1997) 5112--5128,
  \href{http://xxx.lanl.gov/abs/hep-th/9610043}{{\tt hep-th/9610043}}.

\bibitem{Berenstein:2002jq}
D.~E. Berenstein, J.~M. Maldacena, and H.~S. Nastase, ``{Strings in flat space
  and pp waves from N=4 superYang-Mills},'' {\em JHEP} {\bf 0204} (2002) 013,
  \href{http://xxx.lanl.gov/abs/hep-th/0202021}{{\tt hep-th/0202021}}.

\bibitem{Anagnostopoulos:2007fw}
K.~N. Anagnostopoulos, M.~Hanada, J.~Nishimura, and S.~Takeuchi, ``{Monte Carlo
  studies of supersymmetric matrix quantum mechanics with sixteen supercharges
  at finite temperature},'' {\em Phys.Rev.Lett.} {\bf 100} (2008) 021601,
  \href{http://xxx.lanl.gov/abs/0707.4454}{{\tt 0707.4454}}.

\bibitem{Berenstein:2008eg}
D.~E. Berenstein, M.~Hanada, and S.~A. Hartnoll, ``{Multi-matrix models and
  emergent geometry},'' {\em JHEP} {\bf 02} (2009) 010,
  \href{http://xxx.lanl.gov/abs/0805.4658}{{\tt 0805.4658}}.

\bibitem{Asplund:2012tg}
C.~T. Asplund, D.~Berenstein, and E.~Dzienkowski, ``{Large N classical dynamics
  of holographic matrix models},'' {\em Phys. Rev.} {\bf D87} (2013), no.~8,
  084044, \href{http://xxx.lanl.gov/abs/1211.3425}{{\tt 1211.3425}}.

\bibitem{Berenstein:2013tya}
D.~Berenstein and E.~Dzienkowski, ``{Numerical Evidence for Firewalls},''
  \href{http://xxx.lanl.gov/abs/1311.1168}{{\tt 1311.1168}}.

\bibitem{Berenstein:2012ts}
D.~Berenstein and E.~Dzienkowski, ``{Matrix embeddings on flat $R^3$ and the
  geometry of membranes},'' {\em Phys. Rev.} {\bf D86} (2012) 086001,
  \href{http://xxx.lanl.gov/abs/1204.2788}{{\tt 1204.2788}}.

\bibitem{Berenstein:2014pma}
D.~Berenstein, ``{Sketches of emergent geometry in the gauge/gravity
  duality},'' {\em Fortsch. Phys.} {\bf 62} (2014) 776--785,
  \href{http://xxx.lanl.gov/abs/1404.7052}{{\tt 1404.7052}}.

\bibitem{Magan:2016ojb}
J.~M. Magan, ``{Black holes as random particles: entanglement dynamics in
  infinite range and matrix models},''
  \href{http://xxx.lanl.gov/abs/1601.04663}{{\tt 1601.04663}}.
  
\bibitem{connor}
D.~O'Connor and V.~G. Filev, ``Near commuting multi-matrix models,'' {\em JHEP}
  {\bf 04} (2013) 144.

\bibitem{Asplund:2011qj}
C.~Asplund, D.~Berenstein, and D.~Trancanelli, ``{Evidence for fast
  thermalization in the plane-wave matrix model},'' {\em Phys.Rev.Lett.} {\bf
  107} (2011) 171602, \href{http://xxx.lanl.gov/abs/1104.5469}{{\tt
  1104.5469}}. 5 pages, 5 figures, revtex4 format/ v2: minor typos fixed/ v3: 8
  pages, 9 figures, minor changes, includes a supplement as appeared on PRL.

\bibitem{Pramodh:2014jha}
S.~Pramodh and V.~Sahakian, ``{From Black Hole to Qubits: Evidence of Fast
  Scrambling in BMN theory},'' {\em JHEP} {\bf 07} (2015) 067,
  \href{http://xxx.lanl.gov/abs/1412.2396}{{\tt 1412.2396}}.

\bibitem{Seiberg:1997ad}
N.~Seiberg, ``{Why is the matrix model correct?},'' {\em Phys. Rev. Lett.} {\bf
  79} (1997) 3577--3580, \href{http://xxx.lanl.gov/abs/hep-th/9710009}{{\tt
  hep-th/9710009}}.

\bibitem{Bigatti:1997jy}
D.~Bigatti and L.~Susskind, ``{Review of matrix theory},'' in {\em {Strings,
  branes and dualities. Proceedings, NATO Advanced Study Institute, Cargese,
  France, May 26-June 14, 1997}}.
\newblock 1997.
\newblock \href{http://xxx.lanl.gov/abs/hep-th/9712072}{{\tt hep-th/9712072}}.

\bibitem{Taylor:2001vb}
W.~Taylor, ``{M(atrix) theory: Matrix quantum mechanics as a fundamental
  theory},'' {\em Rev. Mod. Phys.} {\bf 73} (2001) 419--462,
  \href{http://xxx.lanl.gov/abs/hep-th/0101126}{{\tt hep-th/0101126}}.

\bibitem{Susskind:1997cw}
L.~Susskind, ``{Another conjecture about M(atrix) theory},''
  \href{http://xxx.lanl.gov/abs/hep-th/9704080}{{\tt hep-th/9704080}}.

\bibitem{Dijkgraaf:1997vv}
R.~Dijkgraaf, E.~Verlinde, and H.~Verlinde, ``Matrix string theory,'' {\em
  Nucl. Phys.} {\bf B500} (1997) 43--61,
  \href{http://xxx.lanl.gov/abs/hep-th/9703030}{{\tt hep-th/9703030}}.

\bibitem{Horowitz:1997fr}
G.~T. Horowitz and E.~J. Martinec, ``Comments on black holes in matrix
  theory,'' {\em Phys. Rev.} {\bf D57} (1998) 4935--4941,
  \href{http://xxx.lanl.gov/abs/hep-th/9710217}{{\tt hep-th/9710217}}.

\bibitem{Maldacena:1999mh}
Maldacena, Juan M. and Russo, Jorge G., ``Large N limit of non-commutative gauge theories,'' {\em JHEP} {\bf 09} (1999) 025,
  \href{http://xxx.lanl.gov/abs/hep-th/9908134}{{\tt hep-th/9908134}}.

\bibitem{Banks:1997hz}
T.~Banks, W.~Fischler, I.~R. Klebanov, and L.~Susskind, ``Schwarzschild black
  holes from matrix theory,'' {\em Phys. Rev. Lett.} {\bf 80} (1998) 226--229,
  \href{http://xxx.lanl.gov/abs/hep-th/9709091}{{\tt hep-th/9709091}}.

\bibitem{Banks:1997tn}
T.~Banks, W.~Fischler, I.~R. Klebanov, and L.~Susskind, ``Schwarzschild black
  holes in matrix theory. ii,'' {\em JHEP} {\bf 01} (1998) 008,
  \href{http://xxx.lanl.gov/abs/hep-th/9711005}{{\tt hep-th/9711005}}.

\bibitem{Maldacena:1997re}
J.~M. Maldacena, ``The large n limit of superconformal field theories and
  supergravity,'' {\em Adv. Theor. Math. Phys.} {\bf 2} (1998) 231--252,
  \href{http://xxx.lanl.gov/abs/hep-th/9711200}{{\tt hep-th/9711200}}.

\bibitem{Witten:1998qj}
E.~Witten, ``Anti-de sitter space and holography,'' {\em Adv. Theor. Math.
  Phys.} {\bf 2} (1998) 253--291,
  \href{http://xxx.lanl.gov/abs/hep-th/9802150}{{\tt hep-th/9802150}}.

\bibitem{Riggins:2012qt}
P.~Riggins and V.~Sahakian, ``{On black hole thermalization, D0 brane dynamics,
  and emergent spacetime},'' {\em Phys.Rev.} {\bf D86} (2012) 046005,
  \href{http://xxx.lanl.gov/abs/1205.3847}{{\tt 1205.3847}}.

\bibitem{Maldacena:2015waa}
J.~Maldacena, S.~H. Shenker, and D.~Stanford, ``{A bound on chaos},''
  \href{http://xxx.lanl.gov/abs/1503.01409}{{\tt 1503.01409}}.

\bibitem{Gur-Ari:2015rcq}
G.~Gur-Ari, M.~Hanada, and S.~H. Shenker, ``{Chaos in Classical D0-Brane
  Mechanics},'' \href{http://xxx.lanl.gov/abs/1512.00019}{{\tt 1512.00019}}.

\bibitem{randommatrix}
P.~Deift and D.~Gioev, {\em Random Matrix Theory: Invariant Ensembles and
  Universality}.
\newblock Courant Institute of Mathematical Sciences, 2000.

\bibitem{Sahakian:1999gj}
V.~Sahakian, ``Black holes and thermodynamics of non-gravitational theories,''
  \href{http://xxx.lanl.gov/abs/hep-th/9906044}{{\tt hep-th/9906044}}.

\bibitem{Martinec:1998ja}
E.~J. Martinec and V.~Sahakian, ``Black holes and the sym phase diagram. ii,''
  {\em Phys. Rev.} {\bf D59} (1999) 124005,
  \href{http://xxx.lanl.gov/abs/hep-th/9810224}{{\tt hep-th/9810224}}.

\bibitem{gottfried}
T.-M.~Yan and Kurt~Gottfried, {\em Quantum Mechanics: Fundamentals}, Springer Science \& Business Media, 2013.

\bibitem{paultrap}
G.~W. F.~G.~Major, Viorica N.~Gheorghe, {\em Charged Particle Traps: Physics
  and Techniques of Charged Particle Field Confinement}, Springer Science \& Business Media, 2006.

\bibitem{hoppe}
J.~Hoppe, ``Quantum theory of a massless relativsitic surface and a
  two-dimensional bound state problem,'' {\em Thesis, MIT} (1982).

\bibitem{Dasgupta:2002hx}
K.~Dasgupta, M.~M. Sheikh-Jabbari, and M.~Van~Raamsdonk, ``{Matrix perturbation
  theory for M theory on a PP wave},'' {\em JHEP} {\bf 0205} (2002) 056,
  \href{http://xxx.lanl.gov/abs/hep-th/0205185}{{\tt hep-th/0205185}}.

\bibitem{Mathur:2005zp}
S.~D. Mathur, ``The fuzzball proposal for black holes: An elementary review,''
  {\em Fortsch. Phys.} {\bf 53} (2005) 793--827,
  \href{http://xxx.lanl.gov/abs/hep-th/0502050}{{\tt hep-th/0502050}}.

\bibitem{Mathur:2008kg}
S.~D. Mathur, ``{Tunneling into fuzzball states},'' {\em Gen.Rel.Grav.} {\bf
  42} (2010) 113--118, \href{http://xxx.lanl.gov/abs/0805.3716}{{\tt
  0805.3716}}.

\bibitem{Bena:2004wv}
I.~Bena, ``Splitting hairs of the three charge black hole,'' {\em Phys. Rev.}
  {\bf D70} (2004) 105018, \href{http://xxx.lanl.gov/abs/hep-th/0404073}{{\tt
  hep-th/0404073}}.

\bibitem{Murugan:2006sn}
A.~Murugan and V.~Sahakian, ``{Emergence of the fuzzy horizon through
  gravitational collapse},'' {\em Phys.Rev.} {\bf D74} (2006) 106010,
  \href{http://xxx.lanl.gov/abs/hep-th/0608103}{{\tt hep-th/0608103}}. 24
  pages, 4 figures/ v2: minor clarifications, citations added.

\bibitem{Almheiri:2012rt}
A.~Almheiri, D.~Marolf, J.~Polchinski, and J.~Sully, ``{Black Holes:
  Complementarity or Firewalls?},'' {\em JHEP} {\bf 1302} (2013) 062,
  \href{http://xxx.lanl.gov/abs/1207.3123}{{\tt 1207.3123}}.

\bibitem{Sahakian:1999bd}
V.~Sahakian, ``Holography, a covariant c-function and the geometry of the
  renormalization group,'' {\em Phys. Rev.} {\bf D62} (2000) 126011,
  \href{http://xxx.lanl.gov/abs/hep-th/9910099}{{\tt hep-th/9910099}}.

\bibitem{Myers:1999ps}
R.~C. Myers, ``Dielectric-branes,'' {\em JHEP} {\bf 12} (1999) 022,
  \href{http://xxx.lanl.gov/abs/hep-th/9910053}{{\tt hep-th/9910053}}.

\bibitem{pescheleisler}
I.~Peschel and V.~Eisler, ``Reduced density matrices and entanglement entropy
  in free lattice models,'' {\em J. Phys. A: Math. Theor.} {\bf 42} (2009)
  504003.

\bibitem{toappear}
V.~Sahakian, ``Holographic c-functions, entanglement entropy, and emergent geometry,'' to appear.

\end{thebibliography}
%
%
%
%

\end{document}